\documentclass[aps,prl,notitlepage,reprint,superscriptaddress,nofootinbib,longbibliography]{revtex4-1}

\usepackage{graphicx}
\usepackage{dcolumn}
\usepackage{bm}
\usepackage{amsmath}
\usepackage{upgreek}
\usepackage[colorlinks,linkcolor=red,citecolor=blue,urlcolor=red]{hyperref}
\usepackage[utf8]{inputenc}
\usepackage[T1]{fontenc}
\usepackage{mathptmx}
\usepackage{ulem}
\usepackage[colorinlistoftodos]{todonotes}
\usepackage[mathlines]{lineno}
\renewcommand{\t}[1]{\mathrm{#1}}
\newcommand{\tHz}[1]{\ensuremath{\mathrm{#1}/\sqrt{\mathrm{Hz}}}}
\AtBeginDocument{%
	\newwrite\bibnotes
	\def\bibnotesext{Notes.bib}
	\immediate\openout\bibnotes=\jobname\bibnotesext
	\immediate\write\bibnotes{@CONTROL{REVTEX41Control}}
	\immediate\write\bibnotes{@CONTROL{%
			apsrev41Control,author="08",editor="1",pages="1",title="0",year="1"}}
	\if@filesw
	\immediate\write\@auxout{\string\citation{apsrev41Control}}%
	\fi
}%

\begin{document}

\title{Sail membranes for optomechanical accelerometry}

\author{A. D. Hyatt}
\affiliation{Wyant College of Optical Sciences, University of Arizona, Tucson, AZ 85721, USA}
\email{atkindavidhyatt@arizona.edu}
	
\author{M. Dey Chowdhury}
\affiliation{Wyant College of Optical Sciences, University of Arizona, Tucson, AZ 85721, USA}


\author{M. Chowdhury}
\affiliation{Mechanical, Automotive and Materials Engineering, University of Windsor, Windsor, ON N9B 3P4,
Canada}

\author{M. J. Ahamed}
\affiliation{Mechanical, Automotive and Materials Engineering, University of Windsor, Windsor, ON N9B 3P4,
Canada}
 
\author{D. J. Wilson}
\email{dalziel@optics.arizona.edu}
\affiliation{Wyant College of Optical Sciences, University of Arizona, Tucson, AZ 85721, USA}

\date{\today}

\begin{abstract}
  Strained membrane resonators have emerged as a promising platform for optomechanical accelerometry; however, the desired combination of low frequency and high $Q$-mass product requires a rethinking of their dissipation dilution engineering.  Applying Bayesian optimization to a Si$_3$N$_4$ membrane, we discover a class of sail-like trampoline resonators in which the frequency is decreased by an order of magnitude while preserving the $Q$-mass product. 
   We demonstrate centimeter-scale sails with kHz frequencies, $Q\sim10^7$ and $Q\times\t{mass}\sim$ 10 g.  Vertically integrating a 7 kHz device with a nanoribbon, we realize a monolithic cavity optomechanical accelerometer with a room temperature thermal noise of $40\;\t{n}g_0/\sqrt{\t{Hz}}$, sufficient to resolve $\mu g_0/\sqrt{\t{Hz}}$ ambient vibration over a bandwidth of 4 kHz with a displacement imprecision of $10^{-14}\;\t{m}/\sqrt{\t{Hz}}$. Cryogenic arrays of sail membranes may be attractive for new physics searches and distributed quantum sensing experiments.
\end{abstract}

\maketitle


Strained membranes have emerged as a key resonator platform for optomechanical force sensing and quantum experiments \cite{metcalfe2014applications}, featuring small mass $m$, high $Q$-frequency ($f$) product, large surface area, and compatibility with micropatterning \cite{tsaturyan2017ultracoherent,norder2024pentagonal} and photonic integration \cite{krause2012high,khokhar2026high}. Over the last decade, a diversity of trampoline, fractal, phononic crystal, and ribbon-like membranes have been studied \cite{engelsen2024ultrahigh}, now routinely achieving $Q>10^8$, $Q\times f>10^{13}$ Hz and thermal force sensitivities of $S_F^\t{th} = 8\pi k_B Tfm/Q\lesssim (10^{-15}\,\tHz{N})^2$ at $f\sim 0.1-1$ MHz and room temperature $T= 300$ K. These advances have been driven by improved understanding of strain-induced dissipation dilution and its geometry dependence \cite{engelsen2024ultrahigh,fedorov2019generalized}, combined with innovative fabrication and design techniques~\cite{ghadimi2018elastic,cupertino2024centimeter,hoj2021ultra,shin2022spiderweb}.

Looking forward, a new generation of fundamental field searches \cite{carney2021mechanical,manley2021searching,liu2021gravitational,tang2025cavity,rousso2026optomechanical} has driven a push towards large-area \cite{cupertino2024centimeter,serra2016microfabrication,martinez2016electromagnetic} and, or, mass-loaded membranes \cite{condos2024ultralow,bawden2025precision,depellette2026strong} with increased $Q\times m$ and reduced $f$, broadly characterized by their thermal acceleration noise \cite{chowdhury2023membrane}
\begin{equation}
	S_a^\t{th} = 8\pi k_B Tf/(Qm).
\end{equation}
While in principle simply increasing the size $L$ of a square membrane can increase acceleration sensitivity as $S_a^\t{th}\propto L^{-4}$ via ``hard-clamped'' dissipation dilution scaling $Q\propto L$ \cite{fedorov2019generalized}, in practice the physical size of strained membranes has been limited to $L\sim \t{cm}$ by wafer-scale fabrication constraints, motivating a search for alternative geometries that increase $Qm/f$.

\begin{figure}[t]
		\vspace{-2mm}
		\includegraphics[width=0.95\columnwidth]{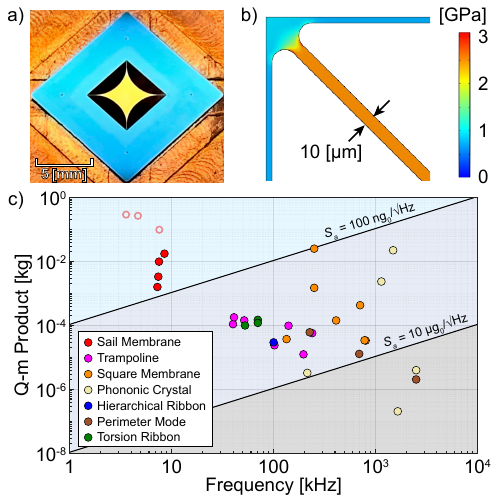}
		\caption{(a) Photo of optimized circular-fillet sail membrane with $2.5\times 2.5\;\t{mm}^2$ pad and 10-$\mu\t{m}$-wide tethers. (b) Simulated von Mises stress near the tether clamps. (c) Compilation of $Q\times m$ versus frequency for trampoline \cite{chowdhury2023membrane,reinhardt2016ultralow,norte2016mechanical,hoj2021ultra}, square \cite{underwood2015measurement,li2024broadband,zwickl2008high,chakram2014dissipation,borrielli2016control}, phononic crystal \cite{cupertino2024centimeter,ghadimi2018elastic,reetz2019analysis,rossi2018measurement,seis2022ground}, hierarchical \cite{bereyhi2022hierarchical}, perimeter mode \cite{bereyhi2022perimeter}, and torsion ribbon \cite{pratt2023nanoscale,hyatt2025ultrahigh} type Si$_3$N$_4$ membranes. Open and solid red circles are simulated and characterized sail membranes described in Figs. \ref{fig:2}-\ref{fig:3}.} 
        \label{fig:1}
		\vspace{-4mm}
	\end{figure}

\begin{figure*}[ht!]
		\includegraphics[width=1.8\columnwidth]{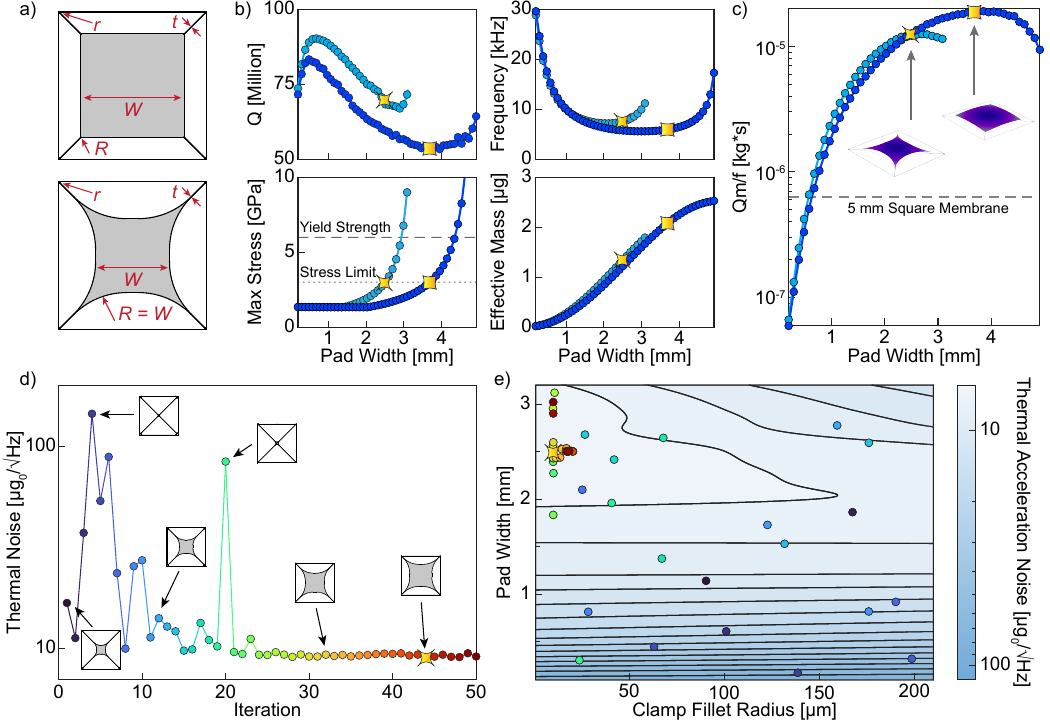}
		\caption{ Optimizing the $Qm/f$ of a Si$_3$N$_4$ trampoline resonator with fixed tether width. (a) Sail-like trampoline geometries with pad-width $W$, tether width $t$, pad fillet radius $R$, and clamp fillet radius $r$. Top: square-fillet sail ($W$ independent of $R$).  Bottom: circular-fillet sail ($W=R$). (b-c) 
        Sweep of $W$ over optimized fillets (yellow markers) with $t=10$ $\mu$m. Plots show quality factor $Q$, frequency $f$, maximum von Mises stress, effective mass $m$, and $Qm/f$ versus $W$. (d-e) Bayesian optimization algorithm convergence (d) and design objective landscape (e) for the bottom device in (a). Colored points in (d) correspond to the same points in (e).}
\label{fig:2}
		\vspace{-2mm}
\end{figure*}

Here we report on a new sail-like membrane resonator that achieves the $Q\times m$ of a square membrane while reducing the fundamental frequency by an order of magnitude---discovered through Bayesian optimization of a strained Si$_3$N$_4$ trampoline. As shown in Fig. \ref{fig:1}, $L\sim\t{cm}$  Si$_3$N$_4$ sails are predicted to have acceleration sensitivities $S_a^\t{th}\sim\;(\t{ng_0/\sqrt{Hz}})^2$ more than an order of magnitude lower than previously achieved with Si$_3$N$_4$ membranes of any geometry.  We explored this by fabricating $L = 0.5$ cm sails, realizing $f<10\;\t{kHz}$, $Q>10^7$, $Q\times m\approx 10$~g, within a factor of 5 of simulation. Vertically integrating a sail with a nanoribbon, we realize a monolithic cavity optomechanical accelerometer \cite{chowdhury2023membrane} with room temperature thermal noise of $40\;\t{n}g_0/\sqrt{\t{Hz}}$, sufficient to resolve $1\;\mu g_0/\sqrt{\t{Hz}}$ ambient vibrations in our laboratory over a 4 kHz bandwidth. Beyond chipscale sensing, cryogenic arrays of sail membranes may be attractive for new physics searches \cite{carney2019ultralight,chowdhury2025optomechanical,rousso2026optomechanical} and distributed quantum sensing experiments \cite{xia2022entanglement,li2026quantum,brady2023entanglement}.

An overview of the resonator geometry and optimization is shown in Fig. \ref{fig:2}. For transverse flexural modes of a membrane with thickness $h$, Young's modulus $E$, and density $\rho$, the quality factor $Q$ and effective mass $m$ depend on the modeshape $u(x,y)$ and stress profile $\sigma(x,y)$, and can be approximated as
\begin{subequations}\begin{align}
    Q & \approx  Q_0\left(1+\frac{\int \sigma  (\nabla u)^2 dA}{Eh^2\int (\nabla^2 u)^2dA}\right)\\
    m & = \rho h\frac{ \int   u^2 dA}{u_\t{max}^2}
\end{align}\end{subequations}
where $Q_0$ is the intrinsic (undiluted) material quality factor and $u_\t{max}$ is the maximum value of $u(x,y)$~\cite{engelsen2024ultrahigh}.

Optimizing the membrane geometry for high $Qm/f$ requires repeated finite-element simulation of candidate designs. Because the stress profile $\sigma(x,y)$ and modeshape $u(x,y)$ must be recomputed for each design, multidimensional parameter sweeps are computationally expensive. We therefore employ Bayesian optimization to efficiently search the design space, adopting a workflow previously described in \cite{hyatt2025ultrahigh}.  

We explore two sail-like trampoline geometries with square and circular fillets (Fig. \ref{fig:2}a), building on the intuition that concentrating mass in the central pad can reduce the resonance frequency \cite{bawden2025precision} while concentrating stress in the clamps can increase $Q$ \cite{bereyhi2019clamp}.  Both designs optimize over the pad width $W$ and clamp fillet radius $r$ but constrain pad fillet radius $R$ differently: the square sail allows $W$, $r$, and $R$ to be optimized independently while the circular fillets fix $R = W$.

Results of a three-parameter $(W,r,R)$ optimization of a square Si$_3$N$_4$ sail with a pre-stress of $1$ GPa, thickness $h = 100\,\t{nm}$, $Q_0 = 60\cdot h/\t{nm}$ \cite{villanueva2014evidence}, tether width $t=10$ $\mu$m \cite{tetherWidth}, and window size $5\times 5\;\t{mm}^2$ are shown in Fig. \ref{fig:2}b-c, revealing the existence of a geometry with $Qm/f$ exceeding that of a square membrane by two orders of magnitude, while maintaining stress in the clamps beneath the material yield strength of $\approx 6\,\t{GPa}$ \cite{norte2016mechanical}. As evident in Fig. \ref{fig:2}b, in which pad size $W$ is swept while fixing $(r,R)$ to the optimal value, this improvement draws from a combination of enhanced $Q$ (established for trampolines and beams \cite{villanueva2014evidence,reinhardt2016ultralow,sadeghi2019influence,bereyhi2019clamp}), mass loading $m\sim W^2$, and a nontrivial $f\sim m^{-1}$ frequency scaling, the latter two saturating as the pad size approaches the window size.

In practice, we favor the circular fillet geometry $(R = W)$ for its relative ease of fabrication and similar optimal performance. (Square fillets are harder to release due to stress gradients; however, a similar geometry as Fig. \ref{fig:4}a was recently realized using a strain-relieved, perforated photonic crystal membrane \cite{norder2026high}.) Shown in Fig. \ref{fig:2}d-e is a Bayesian search for the optimum $(r,W)$  of a $\sigma_0 = 1\,\t{GPa}$, $Q_0 = 60\cdot h/\t{nm}$, $h = 100\,\t{nm}$, $t=10$ $\mu$m \cite{tetherWidth}, $5\times 5$ mm$^2$ wide circular-filleted Si$_3$N$_4$ sail leading to the optimal geometry shown in Fig. \ref{fig:1}. While the two-dimensional $Qm/f$ landscape is non-trivial (contours in Fig. \ref{fig:2}e), the Bayesian search converges to the global optimum after $\sim 20$ iterations, predicting a $(r,W) \approx (15\,\t{\mu\t{m}},\,2.5\,\t{mm})$ geometry with $f\approx 7\,\t{kHz}$, $m\approx 1.3\,\t{\mu}g$, and $Q\approx 70\t{M}$, corresponding to a thermal acceleration noise $S_a^\t{th}\approx 10\,\t{n}g_0/\sqrt{\t{Hz}}$.

To experimentally validate the optimized circular-fillet sail geometry, we fabricated devices using a standard photolithography and wet etch method described in \cite{hyatt2025fabrication}, starting with a 100-nm-thick, double-side-coated Si$_3$N$_4$-on-Si wafer. The centimeter dimensions and strain-concentrated clamps of our devices make them more fragile than typical millimeter-scale trampolines \cite{reinhardt2016ultralow}. We therefore first selectively back-etch chips leaving behind 20-40 $\mu$m of Si before KOH wet-etching to release the membrane (see Methods for details).
To improve yield, tether geometries are further constrained so that the maximum von Mises stress does not exceed 3 GPa. 

\begin{figure}
\includegraphics[width=1\columnwidth]{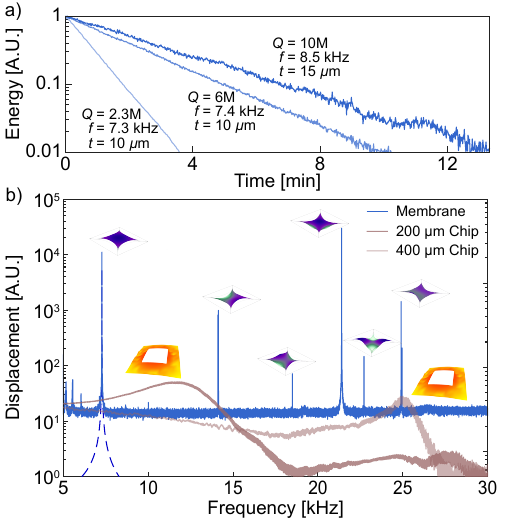}
		\caption{Device characterization. (a) Energy ringdown of three sail membranes with the geometry shown in Fig. \ref{fig:1}a. Exponential fits yield quality factors as high as $Q = 10\times 10^6$.  (b) Free-running vibration spectra (blue) showing the 7 kHz fundamental membrane mode.  Overlaid is the driven response of the 200-$\mu$m-thick device chip (brown) and a similar 400-$\mu$m-thick chip (light brown) taken with a scanning vibrometer at atmospheric pressure, showing broad resonances at 12 and 25 kHz, respectively. Insets: simulated membrane modeshapes and measured chip modeshapes.
        \label{fig:3} }
\end{figure}

Ringdown and free-running vibration measurements of the Fig. \ref{fig:1}a device are shown in Fig. \ref{fig:3}, yielding estimates of mechanical $Q$ and resonance frequency spectrum, respectively. For these measurements, devices were housed in a high-vacuum chamber ($\approx 10^{-8}\;\t{mbar}$) and rested in three-point contact on a metal washer to minimize gas damping and support loss \cite{wilson2009cavity}. A low power $P\sim 0.1 \;\t{mW}$, near-infrared wavelength $\lambda \approx 850$ nm diode laser probed the devices through a viewport. Ringdowns were excited by tapping the vacuum chamber and read out using the optical lever technique \cite{pluchar2025quantum}. 
The vibration spectrum in Fig. \ref{fig:3}b was taken using a double-membrane etalon \cite{chowdhury2023membrane} as described below and in Fig. 4.

We characterized three optimized circular-fillet sails with tether widths $t=\{10,10,15\}\,\mu\t{m}$ \cite{tetherWidth}, and observed fundamental resonance frequencies $f = \{7.3,7.4,8.5\}$ kHz and quality factors $Q = \{2.3,6,10\}\times 10^6$. While the resonance frequencies agree with finite element simulations for a pre-stress $\sigma_0 = 0.9$ GPa, the measured $Q$ factors are lower than predicted by a factor of $5$.  Gas damping is a possible factor; however, the smallest measured damping $f/Q = 0.85$ mHz corresponds to a vacuum pressure of $P_\t{gas}\approx 2\times 10^{-6}\ \t{mbar}$ two orders of magnitude larger than inferred from our ion pump \cite{green2025accurate,reinhardt2024self}.  Another possible explanation is mode-hybridization with the 12-mm-wide device chip, whose fundamental flexure mode frequency is expected to be near that of the membrane \cite{de2022mechanical}. We explored this by conducting a driven response measurement of the chip with a commercial scanning vibrometer \cite{Polytec} (brown data in Fig. \ref{fig:3}[b]), revealing a broad 12 kHz resonance with a displacement profile consistent with simulation and qualitatively overlapping with the membrane mode.

\color{blue} \color{black}

Combined with a simulated effective mass $m=1.3\;\mu$g, the measured damping rates of the circular-fillet devices in Fig. \ref{fig:3}a predict thermal acceleration sensitivities of~$\sqrt{S_a^\t{th}} = $ $\sqrt{8\pi k_B T f/(Qm\beta^2)}\approx \{43,27,22\}\,\t{n} g_0/\sqrt{\t{Hz}}$, where here $\beta= u_\t{max}\int u dA/(\int u^2 dA)\approx 1.2$ is a unitless modal participation factor~\cite{chowdhury2023membrane}. 
By contrast, optimized  $5\times 5\;\t{mm}^2$ square and circular-fillet membranes described in Fig. \ref{fig:2} are predicted to have $\sqrt{S_a^\t{th}} = 7\;\t{n}g_0/\sqrt{\t{Hz}}$ and $10\;\t{n}g_0/\sqrt{\t{Hz}}$, respectively. Simulated and measured devices are compiled in Fig. \ref{fig:1}c alongside previous Si$_3$N$_4$ membrane devices, highlighting access to a novel $\sqrt{S_a^\t{th}}\le 100\;\t{n}g_0/\sqrt{\t{Hz}}$ sensitivity regime.

To explore the enhanced acceleration sensitivity of sail membranes, we fabricated a double-membrane optomechanical accelerometer based on the method described in \cite{chowdhury2023membrane} and carried out vibration measurements as shown in Fig.~\ref{fig:4}. The double-membrane device (Fig. \ref{fig:4}a) consists of a circular-fillet sail with the same dimensions as the Fig. \ref{fig:1}a device vertically-integrated atop a 200-$\mu$m-wide nanoribbon \cite{pratt2023nanoscale} on the same 200-$\mu$m-thick Si chip.  The fundamental flexural mode of the nanoribbon has a resonance frequency of 40 kHz, allowing it to serve as a rigid inertial reference plane \cite{chowdhury2023membrane}.

Vibration measurements were carried out by monitoring the double-membrane transmitted power $P = \bar{P}+\delta P(t)$ and tuning the laser wavelength $\lambda$ to the side of the Fabry-Perot interference fringe $\bar{P}(\lambda)$ (Fig.~\ref{fig:4}b). Classical intensity noise was subtracted by combining transmitted and reference beams on a balanced photodetector \cite{balancedDetCav}. Power spectra were converted to displacement spectra $S_x[\omega] = (\partial \bar{P}/\partial \lambda \cdot \partial \lambda_0/\partial x)^{-2}S_P[\omega]$ using the fitted fringe slope $\partial \bar{P}/\partial \lambda$ and known optomechanical coupling factor $\partial \lambda_0/\partial x = 1/2$. Acceleration spectra $S_a[\omega] = |\chi[\omega]|^{-2} S_x[\omega]$ were inferred by dividing out by the \mbox{mechanical susceptibility, $\chi[\omega]\approx (\omega_0^2 - \omega^2)+i\omega\omega_0/Q$.}

\begin{figure*}[t!]
\includegraphics[width=1.95\columnwidth]{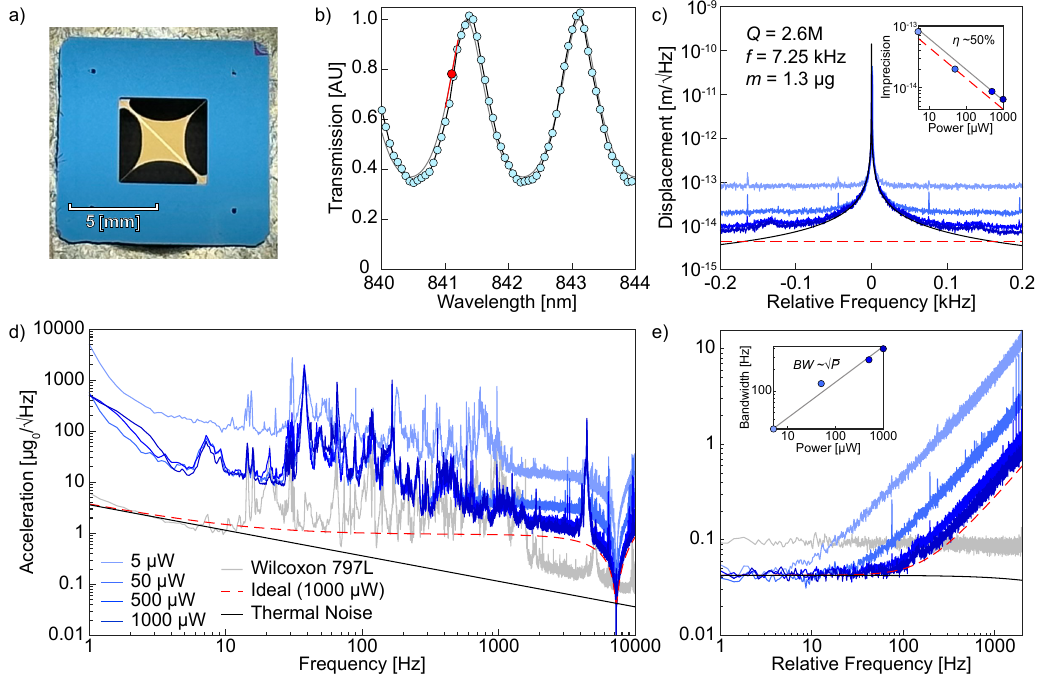}
		\caption{Dual membrane accelerometry with an optimized circular-fillet sail membrane as a test mass and a ribbon membrane as a reference surface. 
        (a) Image of sail-on-ribbon device released from a 200-$\mu$m-thick Si substrate. The top membrane has the sail geometry shown in Fig. \ref{fig:1}a.  The bottom membrane is a 200-$\mu$m-wide ribbon \cite{pratt2023nanoscale}. (b) Transmission versus incident wavelength of double-membrane cavity with finesse $\sim2$. The red point indicates the maximum fringe slope and operating wavelength. (c-e) Free-running displacement (c) and acceleration (d,e) spectra measured with power $\bar{P} = \{5,50,500,1000\}\ \mu\t{W}$. Insets show scaling of noise floor and thermal noise bandwidth versus $\bar{P}$. 
\label{fig:4}}
	\end{figure*}
\color{black}

Displacement and acceleration spectra in Fig. \ref{fig:4}(c-e) were taken with the $f_0\approx 7.3$ kHz, $Q = 2.6\times 10^6$ sail and transmitted powers $\bar{P} = 5$ -$ 1000\,\mu\t{W}$, yielding a displacement imprecision (noise floor) as low as $S_x^\t{imp}\approx (9 \,\t{fm/\sqrt{Hz}})^2$ and acceleration imprecision as low as $S_a^\t{imp} \approx  (40\,\t{n}g_0/\sqrt{\t{Hz}})^2$ over a 3 dB bandwidth of $\Delta f \approx  400\,\t{Hz}$ near mechanical resonance.  The observed power-scaling of $S_x^\t{imp}$ and $\Delta f \approx  (f_0/Q)(S_x^\t{th}/S_x^\t{imp})^{1/2}$ is consistent with shot noise $S_P^\t{imp} = 4\lambda\bar{P}/(hc\eta)$ \cite{balancedDetCav} with a measurement efficiency of $\eta \approx 0.5$ (Fig. \ref{fig:4}[c,e] insets). The acceleration imprecision agrees with the thermal noise prediction $S_a^\t{th}\approx (43\;\t{n}g_0/\sqrt{\t{Hz}})^2$, while baseband noise exceeds the predicted shot-noise-limited value $S_a^\t{imp}(f\ll f_0) \approx S_x^\t{imp}/\omega_0^4\sim (1\;\mu g_0/\sqrt{\t{Hz}})^2$ by over an order of magnitude.  Measurements performed with a commercial seismometer (Wilcoxon 797L, gray curve) rigidly attached to the vacuum chamber suggest that this noise is in excess of seismic and acoustic noise by roughly a factor of 10, pointing to excess laser intensity noise.  

\textit{Summary \& outlook - }In summary, we have used Bayesian optimization to design Si$_3$N$_4$ trampoline resonators with $Qm/f$ products exceeding that of a square membrane by two orders of magnitude, yielding centimeter-scale, microgram devices with $\sim 10$ kHz resonant frequencies and room temperature thermal acceleration sensitivities of $\sim 10\,\t{n}g_0/\sqrt{\t{Hz}}$.

Looking forward, high $Qm/f$ sail membranes hold promise for various applications.  A key motivation for our work is proposed fundamental field sensing experiments based on low noise, scalable membrane accelerometers \cite{chowdhury2023membrane}, including searches for ultralight dark matter \cite{carney2019ultralight,chowdhury2025optomechanical} and high frequency gravitational waves \cite{rousso2026optomechanical}, and study of entanglement-enhanced distributed force sensing protocols \cite{xia2022entanglement,li2026quantum,brady2023entanglement}.  
 Large-area, photonic crystal (PtC)-patterned Si$_3$N$_4$ trampolines \cite{norder2026high}
 are also being used to study radiation pressure effects in next-generation lightsails \cite{norder2026high,michaeli2025direct}. 
The geometry optimization strategy explored here may enable their stiffness to be further reduced; moreover, the flexural modes of such devices have been little studied in vacuum and could give access to rich nonlinear optomechanical effects \cite{michaeli2025optically}.  Finally, membrane-based optomechanical accelerometers \cite{chowdhury2023membrane} have been increasingly explored for practical high speed vibrometry \cite{zhou2021broadband,bawden2025precision,li2024broadband,st2019swept}. The devices we have presented are readily implemented in monolithic \cite {chowdhury2023membrane} or flip-chip cavity approaches \cite{zhou2021broadband,bawden2025precision,li2024broadband}, and integrated with PtC reflectors \cite{norte2016mechanical,agrawal2024focusing}. A particularly attractive approach is to monolithically integrate a PtC sail membrane with a cavity Bragg mirror, building on recent work \cite{khokhar2026high}.

\section*{Methods}
 \textit{Fabrication of sail membranes - }Centimeter-scale $\t{Si_3N_4}$ trampoline resonators have been fabricated using both wet \cite{ghadimi2018elastic,norte2016mechanical} and dry etching \cite{cupertino2024centimeter,norder2026high} techniques; however, several aspects of our device geometries make them uniquely challenging to release: (1) optimizing $Qm/f$ yields localized stress in the tethers beyond that of a typical trampoline resonator; (2) the dimensions of the pad are significantly larger than the tethers and chip thickness, leading to intermediate mass-loading and stress gradients; (3) we desired to open a window through the Si chip without introducing overhang. (The combination of (2) and (3) limits direct use of the SF6 dry-etching technique in \cite{cupertino2024centimeter,norder2026high}.) We overcome these challenges by optimizing sail membranes with constraints on the maximum von Mises stress and 
 adopting a two-stage KOH wet-etching approach:  First, we pre-etch chips from the backside to remove all but $20- 40\;\mu$m of substrate material (deep reactive ion etching may also work for this step); we then perform the double-sided etch and cleaning procedure outlined in \cite{hyatt2025fabrication}. Through trial and error, we found that sail membranes released safely when the stress was limited to 3 GPa. 
\color{black}

\section*{Acknowledgements}
The authors thank Aman Agrawal for helping develop the fabrication procedure. We also thank Osama Jameel and Polytec, Inc (Plymouth, MI, USA) for assistance with vibrometry measurements in Fig. \ref{fig:3}. This work was supported by the National Science Foundation, award no. 2209473 and by the Office of Naval Research MURI, award no. N000142612102. MJA and DJW acknowledge additional support from the Natural Sciences and Engineering Research Council (NSERC) of Canada Alliance International, award no. ALLRP 591060-23.

\bibliography{ref}

\begin{thebibliography}{59}%
\makeatletter
\providecommand \@ifxundefined [1]{%
 \@ifx{#1\undefined}
}%
\providecommand \@ifnum [1]{%
 \ifnum #1\expandafter \@firstoftwo
 \else \expandafter \@secondoftwo
 \fi
}%
\providecommand \@ifx [1]{%
 \ifx #1\expandafter \@firstoftwo
 \else \expandafter \@secondoftwo
 \fi
}%
\providecommand \natexlab [1]{#1}%
\providecommand \enquote  [1]{``#1''}%
\providecommand \bibnamefont  [1]{#1}%
\providecommand \bibfnamefont [1]{#1}%
\providecommand \citenamefont [1]{#1}%
\providecommand \href@noop [0]{\@secondoftwo}%
\providecommand \href [0]{\begingroup \@sanitize@url \@href}%
\providecommand \@href[1]{\@@startlink{#1}\@@href}%
\providecommand \@@href[1]{\endgroup#1\@@endlink}%
\providecommand \@sanitize@url [0]{\catcode `\\12\catcode `\$12\catcode `\&12\catcode `\#12\catcode `\^12\catcode `\_12\catcode `\%12\relax}%
\providecommand \@@startlink[1]{}%
\providecommand \@@endlink[0]{}%
\providecommand \url  [0]{\begingroup\@sanitize@url \@url }%
\providecommand \@url [1]{\endgroup\@href {#1}{\urlprefix }}%
\providecommand \urlprefix  [0]{URL }%
\providecommand \Eprint [0]{\href }%
\providecommand \doibase [0]{http://dx.doi.org/}%
\providecommand \selectlanguage [0]{\@gobble}%
\providecommand \bibinfo  [0]{\@secondoftwo}%
\providecommand \bibfield  [0]{\@secondoftwo}%
\providecommand \translation [1]{[#1]}%
\providecommand \BibitemOpen [0]{}%
\providecommand \bibitemStop [0]{}%
\providecommand \bibitemNoStop [0]{.\EOS\space}%
\providecommand \EOS [0]{\spacefactor3000\relax}%
\providecommand \BibitemShut  [1]{\csname bibitem#1\endcsname}%
\let\auto@bib@innerbib\@empty
\bibitem [{\citenamefont {Metcalfe}(2014)}]{metcalfe2014applications}%
  \BibitemOpen
  \bibfield  {author} {\bibinfo {author} {\bibfnamefont {M.}~\bibnamefont {Metcalfe}},\ }\bibfield  {title} {\enquote {\bibinfo {title} {Applications of cavity optomechanics},}\ }\href {https://pubs.aip.org/aip/apr/article/1/3/031105/123596} {\bibfield  {journal} {\bibinfo  {journal} {Applied Physics Reviews}\ }\textbf {\bibinfo {volume} {1}} (\bibinfo {year} {2014})}\BibitemShut {NoStop}%
\bibitem [{\citenamefont {Tsaturyan}\ \emph {et~al.}(2017)\citenamefont {Tsaturyan}, \citenamefont {Barg}, \citenamefont {Polzik},\ and\ \citenamefont {Schliesser}}]{tsaturyan2017ultracoherent}%
  \BibitemOpen
  \bibfield  {author} {\bibinfo {author} {\bibfnamefont {Y.}~\bibnamefont {Tsaturyan}}, \bibinfo {author} {\bibfnamefont {A.}~\bibnamefont {Barg}}, \bibinfo {author} {\bibfnamefont {E.~S.}\ \bibnamefont {Polzik}}, \ and\ \bibinfo {author} {\bibfnamefont {A.}~\bibnamefont {Schliesser}},\ }\bibfield  {title} {\enquote {\bibinfo {title} {Ultracoherent nanomechanical resonators via soft clamping and dissipation dilution},}\ }\href {https://www.nature.com/articles/nnano.2017.101.pdf} {\bibfield  {journal} {\bibinfo  {journal} {Nat. Nanotechnol.}\ }\textbf {\bibinfo {volume} {12}},\ \bibinfo {pages} {776} (\bibinfo {year} {2017})}\BibitemShut {NoStop}%
\bibitem [{\citenamefont {Norder}\ \emph {et~al.}(2025)\citenamefont {Norder}, \citenamefont {Yin}, \citenamefont {de~Jong}, \citenamefont {Stallone}, \citenamefont {Aydogmus}, \citenamefont {Sberna}, \citenamefont {Bessa},\ and\ \citenamefont {Norte}}]{norder2024pentagonal}%
  \BibitemOpen
  \bibfield  {author} {\bibinfo {author} {\bibfnamefont {L.}~\bibnamefont {Norder}}, \bibinfo {author} {\bibfnamefont {S.}~\bibnamefont {Yin}}, \bibinfo {author} {\bibfnamefont {M.~H.}\ \bibnamefont {de~Jong}}, \bibinfo {author} {\bibfnamefont {F.}~\bibnamefont {Stallone}}, \bibinfo {author} {\bibfnamefont {H.}~\bibnamefont {Aydogmus}}, \bibinfo {author} {\bibfnamefont {P.~M.}\ \bibnamefont {Sberna}}, \bibinfo {author} {\bibfnamefont {M.~A.}\ \bibnamefont {Bessa}}, \ and\ \bibinfo {author} {\bibfnamefont {R.~A.}\ \bibnamefont {Norte}},\ }\bibfield  {title} {\enquote {\bibinfo {title} {Pentagonal photonic crystal mirrors: scalable lightsails with enhanced acceleration via neural topology optimization},}\ }\href {https://doi.org/10.1038/s41467-025-57749-y} {\bibfield  {journal} {\bibinfo  {journal} {Nature Communications}\ }\textbf {\bibinfo {volume} {16}},\ \bibinfo {pages} {2753} (\bibinfo {year} {2025})}\BibitemShut {NoStop}%
\bibitem [{\citenamefont {Krause}\ \emph {et~al.}(2012)\citenamefont {Krause}, \citenamefont {Winger}, \citenamefont {Blasius}, \citenamefont {Lin},\ and\ \citenamefont {Painter}}]{krause2012high}%
  \BibitemOpen
  \bibfield  {author} {\bibinfo {author} {\bibfnamefont {A.~G.}\ \bibnamefont {Krause}}, \bibinfo {author} {\bibfnamefont {M.}~\bibnamefont {Winger}}, \bibinfo {author} {\bibfnamefont {T.~D.}\ \bibnamefont {Blasius}}, \bibinfo {author} {\bibfnamefont {Q.}~\bibnamefont {Lin}}, \ and\ \bibinfo {author} {\bibfnamefont {O.}~\bibnamefont {Painter}},\ }\bibfield  {title} {\enquote {\bibinfo {title} {A high-resolution microchip optomechanical accelerometer},}\ }\href {https://idp.nature.com/authorize/casa?redirect_uri=https://www.nature.com/articles/nphoton.2012.245&casa_token=KX6bdWSa-XoAAAAA:e0fJb4Su915n4ug0sjfHoqUdyZDthBHesbxEJh23fSHrUpbH1CVK_EJIno0ItyEqIM82EfVZPbbFQR49TQ} {\bibfield  {journal} {\bibinfo  {journal} {Nature Photonics}\ }\textbf {\bibinfo {volume} {6}},\ \bibinfo {pages} {768--772} (\bibinfo {year} {2012})}\BibitemShut {NoStop}%
\bibitem [{\citenamefont {Khokhar}\ \emph {et~al.}(2026)\citenamefont {Khokhar}, \citenamefont {Norder}, \citenamefont {Sberna},\ and\ \citenamefont {Norte}}]{khokhar2026high}%
  \BibitemOpen
  \bibfield  {author} {\bibinfo {author} {\bibfnamefont {M.}~\bibnamefont {Khokhar}}, \bibinfo {author} {\bibfnamefont {L.}~\bibnamefont {Norder}}, \bibinfo {author} {\bibfnamefont {P.~M.}\ \bibnamefont {Sberna}}, \ and\ \bibinfo {author} {\bibfnamefont {R.~A.}\ \bibnamefont {Norte}},\ }\bibfield  {title} {\enquote {\bibinfo {title} {High-stress {S}i$_3${N}$_4$ reflective membranes monolithically integrated with cavity {B}ragg mirrors},}\ }\href {https://arxiv.org/pdf/2603.02490} {\bibfield  {journal} {\bibinfo  {journal} {arXiv preprint arXiv:2603.02490}\ } (\bibinfo {year} {2026})}\BibitemShut {NoStop}%
\bibitem [{\citenamefont {Engelsen}\ \emph {et~al.}(2024)\citenamefont {Engelsen}, \citenamefont {Beccari},\ and\ \citenamefont {Kippenberg}}]{engelsen2024ultrahigh}%
  \BibitemOpen
  \bibfield  {author} {\bibinfo {author} {\bibfnamefont {N.~J.}\ \bibnamefont {Engelsen}}, \bibinfo {author} {\bibfnamefont {A.}~\bibnamefont {Beccari}}, \ and\ \bibinfo {author} {\bibfnamefont {T.~J.}\ \bibnamefont {Kippenberg}},\ }\bibfield  {title} {\enquote {\bibinfo {title} {Ultrahigh-quality-factor micro-and nanomechanical resonators using dissipation dilution},}\ }\href {https://doi.org/10.1038/s41565-023-01597-8} {\bibfield  {journal} {\bibinfo  {journal} {Nature Nanotechnology}\ ,\ \bibinfo {pages} {1--13}} (\bibinfo {year} {2024})}\BibitemShut {NoStop}%
\bibitem [{\citenamefont {Fedorov}\ \emph {et~al.}(2019)\citenamefont {Fedorov}, \citenamefont {Engelsen}, \citenamefont {Ghadimi}, \citenamefont {Bereyhi}, \citenamefont {Schilling}, \citenamefont {Wilson},\ and\ \citenamefont {Kippenberg}}]{fedorov2019generalized}%
  \BibitemOpen
  \bibfield  {author} {\bibinfo {author} {\bibfnamefont {S.~A.}\ \bibnamefont {Fedorov}}, \bibinfo {author} {\bibfnamefont {N.~J.}\ \bibnamefont {Engelsen}}, \bibinfo {author} {\bibfnamefont {A.~H.}\ \bibnamefont {Ghadimi}}, \bibinfo {author} {\bibfnamefont {M.~J.}\ \bibnamefont {Bereyhi}}, \bibinfo {author} {\bibfnamefont {R.}~\bibnamefont {Schilling}}, \bibinfo {author} {\bibfnamefont {D.~J.}\ \bibnamefont {Wilson}}, \ and\ \bibinfo {author} {\bibfnamefont {T.~J.}\ \bibnamefont {Kippenberg}},\ }\bibfield  {title} {\enquote {\bibinfo {title} {Generalized dissipation dilution in strained mechanical resonators},}\ }\href {https://doi.org/10.1103/PhysRevB.99.054107} {\bibfield  {journal} {\bibinfo  {journal} {Physical Review B}\ }\textbf {\bibinfo {volume} {99}},\ \bibinfo {pages} {054107} (\bibinfo {year} {2019})}\BibitemShut {NoStop}%
\bibitem [{\citenamefont {Ghadimi}\ \emph {et~al.}(2018)\citenamefont {Ghadimi}, \citenamefont {Fedorov}, \citenamefont {Engelsen}, \citenamefont {Bereyhi}, \citenamefont {Schilling}, \citenamefont {Wilson},\ and\ \citenamefont {Kippenberg}}]{ghadimi2018elastic}%
  \BibitemOpen
  \bibfield  {author} {\bibinfo {author} {\bibfnamefont {A.~H.}\ \bibnamefont {Ghadimi}}, \bibinfo {author} {\bibfnamefont {S.~A.}\ \bibnamefont {Fedorov}}, \bibinfo {author} {\bibfnamefont {N.~J.}\ \bibnamefont {Engelsen}}, \bibinfo {author} {\bibfnamefont {M.~J.}\ \bibnamefont {Bereyhi}}, \bibinfo {author} {\bibfnamefont {R.}~\bibnamefont {Schilling}}, \bibinfo {author} {\bibfnamefont {D.~J.}\ \bibnamefont {Wilson}}, \ and\ \bibinfo {author} {\bibfnamefont {T.~J.}\ \bibnamefont {Kippenberg}},\ }\bibfield  {title} {\enquote {\bibinfo {title} {Elastic strain engineering for ultralow mechanical dissipation},}\ }\href {https://www.science.org/doi/pdf/10.1126/science.aar6939} {\bibfield  {journal} {\bibinfo  {journal} {Science}\ }\textbf {\bibinfo {volume} {360}},\ \bibinfo {pages} {764--768} (\bibinfo {year} {2018})}\BibitemShut {NoStop}%
\bibitem [{\citenamefont {Cupertino}\ \emph {et~al.}(2024)\citenamefont {Cupertino}, \citenamefont {Shin}, \citenamefont {Guo}, \citenamefont {Steeneken}, \citenamefont {Bessa},\ and\ \citenamefont {Norte}}]{cupertino2024centimeter}%
  \BibitemOpen
  \bibfield  {author} {\bibinfo {author} {\bibfnamefont {A.}~\bibnamefont {Cupertino}}, \bibinfo {author} {\bibfnamefont {D.}~\bibnamefont {Shin}}, \bibinfo {author} {\bibfnamefont {L.}~\bibnamefont {Guo}}, \bibinfo {author} {\bibfnamefont {P.~G.}\ \bibnamefont {Steeneken}}, \bibinfo {author} {\bibfnamefont {M.~A.}\ \bibnamefont {Bessa}}, \ and\ \bibinfo {author} {\bibfnamefont {R.~A.}\ \bibnamefont {Norte}},\ }\bibfield  {title} {\enquote {\bibinfo {title} {Centimeter-scale nanomechanical resonators with low dissipation},}\ }\href {https://doi.org/10.1038/s41467-024-48183-7} {\bibfield  {journal} {\bibinfo  {journal} {Nature Communications}\ }\textbf {\bibinfo {volume} {15}},\ \bibinfo {pages} {4255} (\bibinfo {year} {2024})}\BibitemShut {NoStop}%
\bibitem [{\citenamefont {H{\o}j}\ \emph {et~al.}(2021)\citenamefont {H{\o}j}, \citenamefont {Wang}, \citenamefont {Gao}, \citenamefont {Hoff}, \citenamefont {Sigmund},\ and\ \citenamefont {Andersen}}]{hoj2021ultra}%
  \BibitemOpen
  \bibfield  {author} {\bibinfo {author} {\bibfnamefont {D.}~\bibnamefont {H{\o}j}}, \bibinfo {author} {\bibfnamefont {F.}~\bibnamefont {Wang}}, \bibinfo {author} {\bibfnamefont {W.}~\bibnamefont {Gao}}, \bibinfo {author} {\bibfnamefont {U.~B.}\ \bibnamefont {Hoff}}, \bibinfo {author} {\bibfnamefont {O.}~\bibnamefont {Sigmund}}, \ and\ \bibinfo {author} {\bibfnamefont {U.~L.}\ \bibnamefont {Andersen}},\ }\bibfield  {title} {\enquote {\bibinfo {title} {Ultra-coherent nanomechanical resonators based on inverse design},}\ }\href {https://doi.org/10.1038/s41565-023-01597-8} {\bibfield  {journal} {\bibinfo  {journal} {Nature Communications}\ }\textbf {\bibinfo {volume} {12}},\ \bibinfo {pages} {5766} (\bibinfo {year} {2021})}\BibitemShut {NoStop}%
\bibitem [{\citenamefont {Shin}\ \emph {et~al.}(2022)\citenamefont {Shin}, \citenamefont {Cupertino}, \citenamefont {de~Jong}, \citenamefont {Steeneken}, \citenamefont {Bessa},\ and\ \citenamefont {Norte}}]{shin2022spiderweb}%
  \BibitemOpen
  \bibfield  {author} {\bibinfo {author} {\bibfnamefont {D.}~\bibnamefont {Shin}}, \bibinfo {author} {\bibfnamefont {A.}~\bibnamefont {Cupertino}}, \bibinfo {author} {\bibfnamefont {M.~H.}\ \bibnamefont {de~Jong}}, \bibinfo {author} {\bibfnamefont {P.~G.}\ \bibnamefont {Steeneken}}, \bibinfo {author} {\bibfnamefont {M.~A.}\ \bibnamefont {Bessa}}, \ and\ \bibinfo {author} {\bibfnamefont {R.~A.}\ \bibnamefont {Norte}},\ }\bibfield  {title} {\enquote {\bibinfo {title} {Spiderweb nanomechanical resonators via bayesian optimization: inspired by nature and guided by machine learning},}\ }\href {https://advanced.onlinelibrary.wiley.com/doi/pdfdirect/10.1002/adma.202106248?_utm_campaign=mention57529&_utm_content=lnk241316896300&_utm_medium=inline&_utm_source=xakep} {\bibfield  {journal} {\bibinfo  {journal} {Advanced Materials}\ }\textbf {\bibinfo {volume} {34}},\ \bibinfo {pages} {2106248} (\bibinfo {year} {2022})}\BibitemShut {NoStop}%
\bibitem [{\citenamefont {Carney}\ \emph {et~al.}(2021{\natexlab{a}})\citenamefont {Carney}, \citenamefont {Krnjaic}, \citenamefont {Moore}, \citenamefont {Regal}, \citenamefont {Afek}, \citenamefont {Bhave}, \citenamefont {Brubaker}, \citenamefont {Corbitt}, \citenamefont {Cripe}, \citenamefont {Crisosto} \emph {et~al.}}]{carney2021mechanical}%
  \BibitemOpen
  \bibfield  {author} {\bibinfo {author} {\bibfnamefont {D.}~\bibnamefont {Carney}}, \bibinfo {author} {\bibfnamefont {G.}~\bibnamefont {Krnjaic}}, \bibinfo {author} {\bibfnamefont {D.~C.}\ \bibnamefont {Moore}}, \bibinfo {author} {\bibfnamefont {C.~A.}\ \bibnamefont {Regal}}, \bibinfo {author} {\bibfnamefont {G.}~\bibnamefont {Afek}}, \bibinfo {author} {\bibfnamefont {S.}~\bibnamefont {Bhave}}, \bibinfo {author} {\bibfnamefont {B.}~\bibnamefont {Brubaker}}, \bibinfo {author} {\bibfnamefont {T.}~\bibnamefont {Corbitt}}, \bibinfo {author} {\bibfnamefont {J.}~\bibnamefont {Cripe}}, \bibinfo {author} {\bibfnamefont {N.}~\bibnamefont {Crisosto}},  \emph {et~al.},\ }\bibfield  {title} {\enquote {\bibinfo {title} {Mechanical quantum sensing in the search for dark matter},}\ }\href {https://iopscience.iop.org/article/10.1088/2058-9565/abcfcd/pdf} {\bibfield  {journal} {\bibinfo  {journal} {Quantum Science \& Technology}\ }\textbf {\bibinfo {volume} {6}},\ \bibinfo {pages} {024002} (\bibinfo {year}
  {2021}{\natexlab{a}})}\BibitemShut {NoStop}%
\bibitem [{\citenamefont {Manley}\ \emph {et~al.}(2021)\citenamefont {Manley}, \citenamefont {Chowdhury}, \citenamefont {Grin}, \citenamefont {Singh},\ and\ \citenamefont {Wilson}}]{manley2021searching}%
  \BibitemOpen
  \bibfield  {author} {\bibinfo {author} {\bibfnamefont {J.}~\bibnamefont {Manley}}, \bibinfo {author} {\bibfnamefont {M.~D.}\ \bibnamefont {Chowdhury}}, \bibinfo {author} {\bibfnamefont {D.}~\bibnamefont {Grin}}, \bibinfo {author} {\bibfnamefont {S.}~\bibnamefont {Singh}}, \ and\ \bibinfo {author} {\bibfnamefont {D.~J.}\ \bibnamefont {Wilson}},\ }\bibfield  {title} {\enquote {\bibinfo {title} {Searching for vector dark matter with an optomechanical accelerometer},}\ }\href@noop {} {\bibfield  {journal} {\bibinfo  {journal} {Phys. Rev. Lett.}\ }\textbf {\bibinfo {volume} {126}},\ \bibinfo {pages} {061301} (\bibinfo {year} {2021})}\BibitemShut {NoStop}%
\bibitem [{\citenamefont {Liu}\ \emph {et~al.}(2021)\citenamefont {Liu}, \citenamefont {Mummery}, \citenamefont {Zhou},\ and\ \citenamefont {Sillanp{\"a}{\"a}}}]{liu2021gravitational}%
  \BibitemOpen
  \bibfield  {author} {\bibinfo {author} {\bibfnamefont {Y.}~\bibnamefont {Liu}}, \bibinfo {author} {\bibfnamefont {J.}~\bibnamefont {Mummery}}, \bibinfo {author} {\bibfnamefont {J.}~\bibnamefont {Zhou}}, \ and\ \bibinfo {author} {\bibfnamefont {M.~A.}\ \bibnamefont {Sillanp{\"a}{\"a}}},\ }\bibfield  {title} {\enquote {\bibinfo {title} {Gravitational forces between nonclassical mechanical oscillators},}\ }\href {https://journals.aps.org/prapplied/abstract/10.1103/PhysRevApplied.15.034004} {\bibfield  {journal} {\bibinfo  {journal} {Physical Review Applied}\ }\textbf {\bibinfo {volume} {15}},\ \bibinfo {pages} {034004} (\bibinfo {year} {2021})}\BibitemShut {NoStop}%
\bibitem [{\citenamefont {Tang}\ \emph {et~al.}(2025)\citenamefont {Tang}, \citenamefont {Li}, \citenamefont {Sun}, \citenamefont {Cai}, \citenamefont {Li},\ and\ \citenamefont {Liu}}]{tang2025cavity}%
  \BibitemOpen
  \bibfield  {author} {\bibinfo {author} {\bibfnamefont {Z.}~\bibnamefont {Tang}}, \bibinfo {author} {\bibfnamefont {W.}~\bibnamefont {Li}}, \bibinfo {author} {\bibfnamefont {H.}~\bibnamefont {Sun}}, \bibinfo {author} {\bibfnamefont {X.}~\bibnamefont {Cai}}, \bibinfo {author} {\bibfnamefont {T.}~\bibnamefont {Li}}, \ and\ \bibinfo {author} {\bibfnamefont {Y.}~\bibnamefont {Liu}},\ }\bibfield  {title} {\enquote {\bibinfo {title} {Cavity-optomechanical probe of gravity between massive mechanical oscillators},}\ }\href {https://journals.aps.org/pra/abstract/10.1103/m2zy-3ywx} {\bibfield  {journal} {\bibinfo  {journal} {Physical Review A}\ }\textbf {\bibinfo {volume} {112}},\ \bibinfo {pages} {053520} (\bibinfo {year} {2025})}\BibitemShut {NoStop}%
\bibitem [{\citenamefont {Rousso}\ \emph {et~al.}(2026)\citenamefont {Rousso}, \citenamefont {Kunze},\ and\ \citenamefont {Reinhardt}}]{rousso2026optomechanical}%
  \BibitemOpen
  \bibfield  {author} {\bibinfo {author} {\bibfnamefont {D.}~\bibnamefont {Rousso}}, \bibinfo {author} {\bibfnamefont {M.~B.~K.}\ \bibnamefont {Kunze}}, \ and\ \bibinfo {author} {\bibfnamefont {C.}~\bibnamefont {Reinhardt}},\ }\bibfield  {title} {\enquote {\bibinfo {title} {Optomechanical platform for high-frequency gravitational wave and vector dark matter detection},}\ }\href {https://arxiv.org/abs/2601.02576} {\bibfield  {journal} {\bibinfo  {journal} {arXiv preprint arXiv:2601.02576}\ } (\bibinfo {year} {2026})}\BibitemShut {NoStop}%
\bibitem [{\citenamefont {Serra}\ \emph {et~al.}(2016)\citenamefont {Serra}, \citenamefont {Bawaj}, \citenamefont {Borrielli}, \citenamefont {Di~Giuseppe}, \citenamefont {Forte}, \citenamefont {Kralj}, \citenamefont {Malossi}, \citenamefont {Marconi}, \citenamefont {Marin}, \citenamefont {Marino} \emph {et~al.}}]{serra2016microfabrication}%
  \BibitemOpen
  \bibfield  {author} {\bibinfo {author} {\bibfnamefont {E.}~\bibnamefont {Serra}}, \bibinfo {author} {\bibfnamefont {M.}~\bibnamefont {Bawaj}}, \bibinfo {author} {\bibfnamefont {A.}~\bibnamefont {Borrielli}}, \bibinfo {author} {\bibfnamefont {G.}~\bibnamefont {Di~Giuseppe}}, \bibinfo {author} {\bibfnamefont {S.}~\bibnamefont {Forte}}, \bibinfo {author} {\bibfnamefont {N.}~\bibnamefont {Kralj}}, \bibinfo {author} {\bibfnamefont {N.}~\bibnamefont {Malossi}}, \bibinfo {author} {\bibfnamefont {L.}~\bibnamefont {Marconi}}, \bibinfo {author} {\bibfnamefont {F.}~\bibnamefont {Marin}}, \bibinfo {author} {\bibfnamefont {F.}~\bibnamefont {Marino}},  \emph {et~al.},\ }\bibfield  {title} {\enquote {\bibinfo {title} {Microfabrication of large-area circular high-stress silicon nitride membranes for optomechanical applications},}\ }\href@noop {} {\bibfield  {journal} {\bibinfo  {journal} {AIP advances}\ }\textbf {\bibinfo {volume} {6}} (\bibinfo {year} {2016})}\BibitemShut {NoStop}%
\bibitem [{\citenamefont {Martinez}\ \emph {et~al.}(2016)\citenamefont {Martinez}, \citenamefont {Castelli}, \citenamefont {Delmas}, \citenamefont {Sharping},\ and\ \citenamefont {Chiao}}]{martinez2016electromagnetic}%
  \BibitemOpen
  \bibfield  {author} {\bibinfo {author} {\bibfnamefont {L.~A.}\ \bibnamefont {Martinez}}, \bibinfo {author} {\bibfnamefont {A.~R.}\ \bibnamefont {Castelli}}, \bibinfo {author} {\bibfnamefont {W.}~\bibnamefont {Delmas}}, \bibinfo {author} {\bibfnamefont {J.~E.}\ \bibnamefont {Sharping}}, \ and\ \bibinfo {author} {\bibfnamefont {R.}~\bibnamefont {Chiao}},\ }\bibfield  {title} {\enquote {\bibinfo {title} {Electromagnetic coupling to centimeter-scale mechanical membrane resonators via rf cylindrical cavities},}\ }\href {https://pubs.aip.org/aip/adv/article/8/11/115223/127931} {\bibfield  {journal} {\bibinfo  {journal} {New Journal of Physics}\ }\textbf {\bibinfo {volume} {18}},\ \bibinfo {pages} {113015} (\bibinfo {year} {2016})}\BibitemShut {NoStop}%
\bibitem [{\citenamefont {Condos}\ \emph {et~al.}(2025)\citenamefont {Condos}, \citenamefont {Pratt}, \citenamefont {Manley}, \citenamefont {Agrawal}, \citenamefont {Schlamminger}, \citenamefont {Pluchar},\ and\ \citenamefont {Wilson}}]{condos2024ultralow}%
  \BibitemOpen
  \bibfield  {author} {\bibinfo {author} {\bibfnamefont {C.}~\bibnamefont {Condos}}, \bibinfo {author} {\bibfnamefont {J.}~\bibnamefont {Pratt}}, \bibinfo {author} {\bibfnamefont {J.}~\bibnamefont {Manley}}, \bibinfo {author} {\bibfnamefont {A.}~\bibnamefont {Agrawal}}, \bibinfo {author} {\bibfnamefont {S.}~\bibnamefont {Schlamminger}}, \bibinfo {author} {\bibfnamefont {C.}~\bibnamefont {Pluchar}}, \ and\ \bibinfo {author} {\bibfnamefont {D.}~\bibnamefont {Wilson}},\ }\bibfield  {title} {\enquote {\bibinfo {title} {Ultralow loss torsion micropendula for chipscale gravimetry},}\ }\href {https://doi.org/10.1103/nmx5-hygh} {\bibfield  {journal} {\bibinfo  {journal} {Physical Review Letters}\ }\textbf {\bibinfo {volume} {134}},\ \bibinfo {pages} {253602} (\bibinfo {year} {2025})}\BibitemShut {NoStop}%
\bibitem [{\citenamefont {Bawden}\ \emph {et~al.}(2025)\citenamefont {Bawden}, \citenamefont {Carey}, \citenamefont {Yeo}, \citenamefont {Arora}, \citenamefont {Sementilli}, \citenamefont {Valenzuela}, \citenamefont {Romero}, \citenamefont {Harris}, \citenamefont {Wegener},\ and\ \citenamefont {Bowen}}]{bawden2025precision}%
  \BibitemOpen
  \bibfield  {author} {\bibinfo {author} {\bibfnamefont {N.}~\bibnamefont {Bawden}}, \bibinfo {author} {\bibfnamefont {B.~J.}\ \bibnamefont {Carey}}, \bibinfo {author} {\bibfnamefont {P.-M.}\ \bibnamefont {Yeo}}, \bibinfo {author} {\bibfnamefont {N.}~\bibnamefont {Arora}}, \bibinfo {author} {\bibfnamefont {L.}~\bibnamefont {Sementilli}}, \bibinfo {author} {\bibfnamefont {V.~M.}\ \bibnamefont {Valenzuela}}, \bibinfo {author} {\bibfnamefont {E.}~\bibnamefont {Romero}}, \bibinfo {author} {\bibfnamefont {G.~I.}\ \bibnamefont {Harris}}, \bibinfo {author} {\bibfnamefont {M.}~\bibnamefont {Wegener}}, \ and\ \bibinfo {author} {\bibfnamefont {W.~P.}\ \bibnamefont {Bowen}},\ }\bibfield  {title} {\enquote {\bibinfo {title} {Precision optomechanical accelerometer via hybrid test-mass integration},}\ }\href {https://link.aps.org/doi/10.1103/knpw-1mdj} {\bibfield  {journal} {\bibinfo  {journal} {Physical Review Applied}\ }\textbf {\bibinfo {volume} {24}},\ \bibinfo {pages} {064008} (\bibinfo {year} {2025})}\BibitemShut
  {NoStop}%
\bibitem [{\citenamefont {Depellette}\ \emph {et~al.}(2026)\citenamefont {Depellette}, \citenamefont {Rej}, \citenamefont {Cutting},\ and\ \citenamefont {Sillanp{\"a}{\"a}}}]{depellette2026strong}%
  \BibitemOpen
  \bibfield  {author} {\bibinfo {author} {\bibfnamefont {J.}~\bibnamefont {Depellette}}, \bibinfo {author} {\bibfnamefont {E.}~\bibnamefont {Rej}}, \bibinfo {author} {\bibfnamefont {R.}~\bibnamefont {Cutting}}, \ and\ \bibinfo {author} {\bibfnamefont {M.~A.}\ \bibnamefont {Sillanp{\"a}{\"a}}},\ }\bibfield  {title} {\enquote {\bibinfo {title} {Strong actuation of mass-loaded membranes for gravity studies at the milligram scale},}\ }\href {https://pubs.aip.org/aip/jap/article/139/14/144501/3386861} {\bibfield  {journal} {\bibinfo  {journal} {Journal of Applied Physics}\ }\textbf {\bibinfo {volume} {139}} (\bibinfo {year} {2026})}\BibitemShut {NoStop}%
\bibitem [{\citenamefont {Chowdhury}\ \emph {et~al.}(2023)\citenamefont {Chowdhury}, \citenamefont {Agrawal},\ and\ \citenamefont {Wilson}}]{chowdhury2023membrane}%
  \BibitemOpen
  \bibfield  {author} {\bibinfo {author} {\bibfnamefont {M.~D.}\ \bibnamefont {Chowdhury}}, \bibinfo {author} {\bibfnamefont {A.~R.}\ \bibnamefont {Agrawal}}, \ and\ \bibinfo {author} {\bibfnamefont {D.~J.}\ \bibnamefont {Wilson}},\ }\bibfield  {title} {\enquote {\bibinfo {title} {Membrane-based optomechanical accelerometry},}\ }\href {https://link.aps.org/pdf/10.1103/PhysRevApplied.19.024011?casa_token=N-268TjmfFkAAAAA:WJyyjoj4LN-CwmEiLtbUuVNP_II12HQGXk-TUcpSCwfmE1pb7-9MT-Q6g2WwU9Xg6ng_NLBVzb8OOxk} {\bibfield  {journal} {\bibinfo  {journal} {Physical Review Applied}\ }\textbf {\bibinfo {volume} {19}},\ \bibinfo {pages} {024011} (\bibinfo {year} {2023})}\BibitemShut {NoStop}%
\bibitem [{\citenamefont {Reinhardt}\ \emph {et~al.}(2016)\citenamefont {Reinhardt}, \citenamefont {M{\"u}ller}, \citenamefont {Bourassa},\ and\ \citenamefont {Sankey}}]{reinhardt2016ultralow}%
  \BibitemOpen
  \bibfield  {author} {\bibinfo {author} {\bibfnamefont {C.}~\bibnamefont {Reinhardt}}, \bibinfo {author} {\bibfnamefont {T.}~\bibnamefont {M{\"u}ller}}, \bibinfo {author} {\bibfnamefont {A.}~\bibnamefont {Bourassa}}, \ and\ \bibinfo {author} {\bibfnamefont {J.~C.}\ \bibnamefont {Sankey}},\ }\bibfield  {title} {\enquote {\bibinfo {title} {Ultralow-noise sin trampoline resonators for sensing and optomechanics},}\ }\href {https://link.aps.org/pdf/10.1103/PhysRevX.6.021001} {\bibfield  {journal} {\bibinfo  {journal} {Physical Review X}\ }\textbf {\bibinfo {volume} {6}},\ \bibinfo {pages} {021001} (\bibinfo {year} {2016})}\BibitemShut {NoStop}%
\bibitem [{\citenamefont {Norte}\ \emph {et~al.}(2016)\citenamefont {Norte}, \citenamefont {Moura},\ and\ \citenamefont {Gr{\"o}blacher}}]{norte2016mechanical}%
  \BibitemOpen
  \bibfield  {author} {\bibinfo {author} {\bibfnamefont {R.~A.}\ \bibnamefont {Norte}}, \bibinfo {author} {\bibfnamefont {J.~P.}\ \bibnamefont {Moura}}, \ and\ \bibinfo {author} {\bibfnamefont {S.}~\bibnamefont {Gr{\"o}blacher}},\ }\bibfield  {title} {\enquote {\bibinfo {title} {Mechanical resonators for quantum optomechanics experiments at room temperature},}\ }\href {https://journals.aps.org/prl/abstract/10.1103/PhysRevLett.116.147202} {\bibfield  {journal} {\bibinfo  {journal} {Physical Review Letters}\ }\textbf {\bibinfo {volume} {116}},\ \bibinfo {pages} {147202} (\bibinfo {year} {2016})}\BibitemShut {NoStop}%
\bibitem [{\citenamefont {Underwood}\ \emph {et~al.}(2015)\citenamefont {Underwood}, \citenamefont {Mason}, \citenamefont {Lee}, \citenamefont {Xu}, \citenamefont {Jiang}, \citenamefont {Shkarin}, \citenamefont {B{\o}rkje}, \citenamefont {Girvin},\ and\ \citenamefont {Harris}}]{underwood2015measurement}%
  \BibitemOpen
  \bibfield  {author} {\bibinfo {author} {\bibfnamefont {M.}~\bibnamefont {Underwood}}, \bibinfo {author} {\bibfnamefont {D.}~\bibnamefont {Mason}}, \bibinfo {author} {\bibfnamefont {D.}~\bibnamefont {Lee}}, \bibinfo {author} {\bibfnamefont {H.}~\bibnamefont {Xu}}, \bibinfo {author} {\bibfnamefont {L.}~\bibnamefont {Jiang}}, \bibinfo {author} {\bibfnamefont {A.}~\bibnamefont {Shkarin}}, \bibinfo {author} {\bibfnamefont {K.}~\bibnamefont {B{\o}rkje}}, \bibinfo {author} {\bibfnamefont {S.}~\bibnamefont {Girvin}}, \ and\ \bibinfo {author} {\bibfnamefont {J.}~\bibnamefont {Harris}},\ }\bibfield  {title} {\enquote {\bibinfo {title} {Measurement of the motional sidebands of a nanogram-scale oscillator in the quantum regime},}\ }\href {https://journals.aps.org/pra/abstract/10.1103/PhysRevA.92.061801} {\bibfield  {journal} {\bibinfo  {journal} {Physical Review A}\ }\textbf {\bibinfo {volume} {92}},\ \bibinfo {pages} {061801} (\bibinfo {year} {2015})}\BibitemShut {NoStop}%
\bibitem [{\citenamefont {Li}\ \emph {et~al.}(2024)\citenamefont {Li}, \citenamefont {Liu}, \citenamefont {Liu}, \citenamefont {Gu}, \citenamefont {Liu}, \citenamefont {Zhou}, \citenamefont {Xing}, \citenamefont {Shi}, \citenamefont {Tang},\ and\ \citenamefont {Liu}}]{li2024broadband}%
  \BibitemOpen
  \bibfield  {author} {\bibinfo {author} {\bibfnamefont {W.}~\bibnamefont {Li}}, \bibinfo {author} {\bibfnamefont {W.}~\bibnamefont {Liu}}, \bibinfo {author} {\bibfnamefont {C.}~\bibnamefont {Liu}}, \bibinfo {author} {\bibfnamefont {Y.}~\bibnamefont {Gu}}, \bibinfo {author} {\bibfnamefont {L.}~\bibnamefont {Liu}}, \bibinfo {author} {\bibfnamefont {Y.}~\bibnamefont {Zhou}}, \bibinfo {author} {\bibfnamefont {E.}~\bibnamefont {Xing}}, \bibinfo {author} {\bibfnamefont {Y.}~\bibnamefont {Shi}}, \bibinfo {author} {\bibfnamefont {J.}~\bibnamefont {Tang}}, \ and\ \bibinfo {author} {\bibfnamefont {J.}~\bibnamefont {Liu}},\ }\bibfield  {title} {\enquote {\bibinfo {title} {Broadband optomechanical accelerometer reaching the thermomechanical limit based on suspended si 3 n 4 membrane resonator},}\ }\href {https://ieeexplore.ieee.org/abstract/document/10507739?casa_token=IANrWrTs5esAAAAA:2QxrUbG4GuDxUIRd7C3zaYWiGepziUSeAYco0QYSjdmqdEzyAJiKTlXLlLyOENlvgHr_nPzm1w} {\bibfield  {journal} {\bibinfo  {journal} {IEEE Sensors
  Journal}\ }\textbf {\bibinfo {volume} {24}},\ \bibinfo {pages} {17528--17536} (\bibinfo {year} {2024})}\BibitemShut {NoStop}%
\bibitem [{\citenamefont {Zwickl}\ \emph {et~al.}(2008)\citenamefont {Zwickl}, \citenamefont {Shanks}, \citenamefont {Jayich}, \citenamefont {Yang}, \citenamefont {Bleszynski~Jayich}, \citenamefont {Thompson},\ and\ \citenamefont {Harris}}]{zwickl2008high}%
  \BibitemOpen
  \bibfield  {author} {\bibinfo {author} {\bibfnamefont {B.}~\bibnamefont {Zwickl}}, \bibinfo {author} {\bibfnamefont {W.}~\bibnamefont {Shanks}}, \bibinfo {author} {\bibfnamefont {A.}~\bibnamefont {Jayich}}, \bibinfo {author} {\bibfnamefont {C.}~\bibnamefont {Yang}}, \bibinfo {author} {\bibfnamefont {A.}~\bibnamefont {Bleszynski~Jayich}}, \bibinfo {author} {\bibfnamefont {J.}~\bibnamefont {Thompson}}, \ and\ \bibinfo {author} {\bibfnamefont {J.}~\bibnamefont {Harris}},\ }\bibfield  {title} {\enquote {\bibinfo {title} {High quality mechanical and optical properties of commercial silicon nitride membranes},}\ }\href {https://pubs.aip.org/aip/apl/article/92/10/103125/145524} {\bibfield  {journal} {\bibinfo  {journal} {Applied Physics Letters}\ }\textbf {\bibinfo {volume} {92}} (\bibinfo {year} {2008})}\BibitemShut {NoStop}%
\bibitem [{\citenamefont {Chakram}\ \emph {et~al.}(2014)\citenamefont {Chakram}, \citenamefont {Patil}, \citenamefont {Chang},\ and\ \citenamefont {Vengalattore}}]{chakram2014dissipation}%
  \BibitemOpen
  \bibfield  {author} {\bibinfo {author} {\bibfnamefont {S.}~\bibnamefont {Chakram}}, \bibinfo {author} {\bibfnamefont {Y.}~\bibnamefont {Patil}}, \bibinfo {author} {\bibfnamefont {L.}~\bibnamefont {Chang}}, \ and\ \bibinfo {author} {\bibfnamefont {M.}~\bibnamefont {Vengalattore}},\ }\bibfield  {title} {\enquote {\bibinfo {title} {Dissipation in ultrahigh quality factor sin membrane resonators},}\ }\href {https://doi.org/10.1103/PhysRevLett.112.127201} {\bibfield  {journal} {\bibinfo  {journal} {Physical Review Letters}\ }\textbf {\bibinfo {volume} {112}},\ \bibinfo {pages} {127201} (\bibinfo {year} {2014})}\BibitemShut {NoStop}%
\bibitem [{\citenamefont {Borrielli}\ \emph {et~al.}(2016)\citenamefont {Borrielli}, \citenamefont {Marconi}, \citenamefont {Marin}, \citenamefont {Marino}, \citenamefont {Morana}, \citenamefont {Pandraud}, \citenamefont {Pontin}, \citenamefont {Prodi}, \citenamefont {Sarro}, \citenamefont {Serra},\ and\ \citenamefont {Bonaldi}}]{borrielli2016control}%
  \BibitemOpen
  \bibfield  {author} {\bibinfo {author} {\bibfnamefont {A.}~\bibnamefont {Borrielli}}, \bibinfo {author} {\bibfnamefont {L.}~\bibnamefont {Marconi}}, \bibinfo {author} {\bibfnamefont {F.}~\bibnamefont {Marin}}, \bibinfo {author} {\bibfnamefont {F.}~\bibnamefont {Marino}}, \bibinfo {author} {\bibfnamefont {B.}~\bibnamefont {Morana}}, \bibinfo {author} {\bibfnamefont {G.}~\bibnamefont {Pandraud}}, \bibinfo {author} {\bibfnamefont {A.}~\bibnamefont {Pontin}}, \bibinfo {author} {\bibfnamefont {G.~A.}\ \bibnamefont {Prodi}}, \bibinfo {author} {\bibfnamefont {P.~M.}\ \bibnamefont {Sarro}}, \bibinfo {author} {\bibfnamefont {E.}~\bibnamefont {Serra}}, \ and\ \bibinfo {author} {\bibfnamefont {M.}~\bibnamefont {Bonaldi}},\ }\bibfield  {title} {\enquote {\bibinfo {title} {Control of recoil losses in nanomechanical sin membrane resonators},}\ }\href {\doibase 10.1103/PhysRevB.94.121403} {\bibfield  {journal} {\bibinfo  {journal} {Physical Review B}\ }\textbf {\bibinfo {volume} {94}},\ \bibinfo {pages} {121403(R)}
  (\bibinfo {year} {2016})}\BibitemShut {NoStop}%
\bibitem [{\citenamefont {Reetz}\ \emph {et~al.}(2019)\citenamefont {Reetz}, \citenamefont {Fischer}, \citenamefont {Assumpcao}, \citenamefont {McNally}, \citenamefont {Burns}, \citenamefont {Sankey},\ and\ \citenamefont {Regal}}]{reetz2019analysis}%
  \BibitemOpen
  \bibfield  {author} {\bibinfo {author} {\bibfnamefont {C.}~\bibnamefont {Reetz}}, \bibinfo {author} {\bibfnamefont {R.}~\bibnamefont {Fischer}}, \bibinfo {author} {\bibfnamefont {G.~G.}\ \bibnamefont {Assumpcao}}, \bibinfo {author} {\bibfnamefont {D.~P.}\ \bibnamefont {McNally}}, \bibinfo {author} {\bibfnamefont {P.~S.}\ \bibnamefont {Burns}}, \bibinfo {author} {\bibfnamefont {J.~C.}\ \bibnamefont {Sankey}}, \ and\ \bibinfo {author} {\bibfnamefont {C.~A.}\ \bibnamefont {Regal}},\ }\bibfield  {title} {\enquote {\bibinfo {title} {Analysis of membrane phononic crystals with wide band gaps and low-mass defects},}\ }\href {https://journals.aps.org/prapplied/abstract/10.1103/PhysRevApplied.12.044027} {\bibfield  {journal} {\bibinfo  {journal} {Physical Review Applied}\ }\textbf {\bibinfo {volume} {12}},\ \bibinfo {pages} {044027} (\bibinfo {year} {2019})}\BibitemShut {NoStop}%
\bibitem [{\citenamefont {Rossi}\ \emph {et~al.}(2018)\citenamefont {Rossi}, \citenamefont {Mason}, \citenamefont {Chen}, \citenamefont {Tsaturyan},\ and\ \citenamefont {Schliesser}}]{rossi2018measurement}%
  \BibitemOpen
  \bibfield  {author} {\bibinfo {author} {\bibfnamefont {M.}~\bibnamefont {Rossi}}, \bibinfo {author} {\bibfnamefont {D.}~\bibnamefont {Mason}}, \bibinfo {author} {\bibfnamefont {J.}~\bibnamefont {Chen}}, \bibinfo {author} {\bibfnamefont {Y.}~\bibnamefont {Tsaturyan}}, \ and\ \bibinfo {author} {\bibfnamefont {A.}~\bibnamefont {Schliesser}},\ }\bibfield  {title} {\enquote {\bibinfo {title} {Measurement-based quantum control of mechanical motion},}\ }\href {https://idp.nature.com/authorize/casa?redirect_uri=https://www.nature.com/articles/s41586-018-0643-8&casa_token=RSrP0KbP_KgAAAAA:0XW5F_2-41EGmJ3SU7d1DtEcpWRRiSoQtMzmzmVGo2fy2GbtBDYmimPa2q8q5fNj74QICaj-KOxk6aZA1w} {\bibfield  {journal} {\bibinfo  {journal} {Nature}\ }\textbf {\bibinfo {volume} {563}},\ \bibinfo {pages} {53--58} (\bibinfo {year} {2018})}\BibitemShut {NoStop}%
\bibitem [{\citenamefont {Seis}\ \emph {et~al.}(2022)\citenamefont {Seis}, \citenamefont {Capelle}, \citenamefont {Langman}, \citenamefont {Saarinen}, \citenamefont {Planz},\ and\ \citenamefont {Schliesser}}]{seis2022ground}%
  \BibitemOpen
  \bibfield  {author} {\bibinfo {author} {\bibfnamefont {Y.}~\bibnamefont {Seis}}, \bibinfo {author} {\bibfnamefont {T.}~\bibnamefont {Capelle}}, \bibinfo {author} {\bibfnamefont {E.}~\bibnamefont {Langman}}, \bibinfo {author} {\bibfnamefont {S.}~\bibnamefont {Saarinen}}, \bibinfo {author} {\bibfnamefont {E.}~\bibnamefont {Planz}}, \ and\ \bibinfo {author} {\bibfnamefont {A.}~\bibnamefont {Schliesser}},\ }\bibfield  {title} {\enquote {\bibinfo {title} {Ground state cooling of an ultracoherent electromechanical system},}\ }\href {https://www.nature.com/articles/s41467-022-29115-9.pdf} {\bibfield  {journal} {\bibinfo  {journal} {Nature Communications}\ }\textbf {\bibinfo {volume} {13}},\ \bibinfo {pages} {1507} (\bibinfo {year} {2022})}\BibitemShut {NoStop}%
\bibitem [{\citenamefont {Bereyhi}\ \emph {et~al.}(2022{\natexlab{a}})\citenamefont {Bereyhi}, \citenamefont {Beccari}, \citenamefont {Groth}, \citenamefont {Fedorov}, \citenamefont {Arabmoheghi}, \citenamefont {Kippenberg},\ and\ \citenamefont {Engelsen}}]{bereyhi2022hierarchical}%
  \BibitemOpen
  \bibfield  {author} {\bibinfo {author} {\bibfnamefont {M.~J.}\ \bibnamefont {Bereyhi}}, \bibinfo {author} {\bibfnamefont {A.}~\bibnamefont {Beccari}}, \bibinfo {author} {\bibfnamefont {R.}~\bibnamefont {Groth}}, \bibinfo {author} {\bibfnamefont {S.~A.}\ \bibnamefont {Fedorov}}, \bibinfo {author} {\bibfnamefont {A.}~\bibnamefont {Arabmoheghi}}, \bibinfo {author} {\bibfnamefont {T.~J.}\ \bibnamefont {Kippenberg}}, \ and\ \bibinfo {author} {\bibfnamefont {N.~J.}\ \bibnamefont {Engelsen}},\ }\bibfield  {title} {\enquote {\bibinfo {title} {Hierarchical tensile structures with ultralow mechanical dissipation},}\ }\href {https://doi.org/10.1038/s41467-022-30586-z} {\bibfield  {journal} {\bibinfo  {journal} {Nature Communications}\ }\textbf {\bibinfo {volume} {13}},\ \bibinfo {pages} {3097} (\bibinfo {year} {2022}{\natexlab{a}})}\BibitemShut {NoStop}%
\bibitem [{\citenamefont {Bereyhi}\ \emph {et~al.}(2022{\natexlab{b}})\citenamefont {Bereyhi}, \citenamefont {Arabmoheghi}, \citenamefont {Beccari}, \citenamefont {Fedorov}, \citenamefont {Huang}, \citenamefont {Kippenberg},\ and\ \citenamefont {Engelsen}}]{bereyhi2022perimeter}%
  \BibitemOpen
  \bibfield  {author} {\bibinfo {author} {\bibfnamefont {M.~J.}\ \bibnamefont {Bereyhi}}, \bibinfo {author} {\bibfnamefont {A.}~\bibnamefont {Arabmoheghi}}, \bibinfo {author} {\bibfnamefont {A.}~\bibnamefont {Beccari}}, \bibinfo {author} {\bibfnamefont {S.~A.}\ \bibnamefont {Fedorov}}, \bibinfo {author} {\bibfnamefont {G.}~\bibnamefont {Huang}}, \bibinfo {author} {\bibfnamefont {T.~J.}\ \bibnamefont {Kippenberg}}, \ and\ \bibinfo {author} {\bibfnamefont {N.~J.}\ \bibnamefont {Engelsen}},\ }\bibfield  {title} {\enquote {\bibinfo {title} {Perimeter modes of nanomechanical resonators exhibit quality factors exceeding 10 9 at room temperature},}\ }\href {https://link.aps.org/pdf/10.1103/PhysRevX.12.021036} {\bibfield  {journal} {\bibinfo  {journal} {Physical Review X}\ }\textbf {\bibinfo {volume} {12}},\ \bibinfo {pages} {021036} (\bibinfo {year} {2022}{\natexlab{b}})}\BibitemShut {NoStop}%
\bibitem [{\citenamefont {Pratt}\ \emph {et~al.}(2023)\citenamefont {Pratt}, \citenamefont {Agrawal}, \citenamefont {Condos}, \citenamefont {Pluchar}, \citenamefont {Schlamminger},\ and\ \citenamefont {Wilson}}]{pratt2023nanoscale}%
  \BibitemOpen
  \bibfield  {author} {\bibinfo {author} {\bibfnamefont {J.~R.}\ \bibnamefont {Pratt}}, \bibinfo {author} {\bibfnamefont {A.~R.}\ \bibnamefont {Agrawal}}, \bibinfo {author} {\bibfnamefont {C.~A.}\ \bibnamefont {Condos}}, \bibinfo {author} {\bibfnamefont {C.~M.}\ \bibnamefont {Pluchar}}, \bibinfo {author} {\bibfnamefont {S.}~\bibnamefont {Schlamminger}}, \ and\ \bibinfo {author} {\bibfnamefont {D.~J.}\ \bibnamefont {Wilson}},\ }\bibfield  {title} {\enquote {\bibinfo {title} {Nanoscale torsional dissipation dilution for quantum experiments and precision measurement},}\ }\href {https://journals.aps.org/prx/abstract/10.1103/PhysRevX.13.011018} {\bibfield  {journal} {\bibinfo  {journal} {Phys. Rev. X}\ }\textbf {\bibinfo {volume} {13}},\ \bibinfo {pages} {011018} (\bibinfo {year} {2023})}\BibitemShut {NoStop}%
\bibitem [{\citenamefont {Hyatt}\ \emph {et~al.}(2025{\natexlab{a}})\citenamefont {Hyatt}, \citenamefont {Agrawal}, \citenamefont {Pluchar}, \citenamefont {Condos},\ and\ \citenamefont {Wilson}}]{hyatt2025ultrahigh}%
  \BibitemOpen
  \bibfield  {author} {\bibinfo {author} {\bibfnamefont {A.~D.}\ \bibnamefont {Hyatt}}, \bibinfo {author} {\bibfnamefont {A.~R.}\ \bibnamefont {Agrawal}}, \bibinfo {author} {\bibfnamefont {C.~M.}\ \bibnamefont {Pluchar}}, \bibinfo {author} {\bibfnamefont {C.~A.}\ \bibnamefont {Condos}}, \ and\ \bibinfo {author} {\bibfnamefont {D.~J.}\ \bibnamefont {Wilson}},\ }\bibfield  {title} {\enquote {\bibinfo {title} {Ultrahigh-q torsional nanomechanics through bayesian optimization},}\ }\href {https://pubs.acs.org/doi/pdf/10.1021/acs.nanolett.5c06306?casa_token=lORIGoEhvdsAAAAA:mc44UQw53O-IGDWE1odJjs-PRmtQ2wtvrLZTxUebZc-S5p7Br5FUWpT95QoDK32PSj7RMEqTsib_3NQv} {\bibfield  {journal} {\bibinfo  {journal} {Nano Letters}\ } (\bibinfo {year} {2025}{\natexlab{a}})}\BibitemShut {NoStop}%
\bibitem [{\citenamefont {Carney}\ \emph {et~al.}(2021{\natexlab{b}})\citenamefont {Carney}, \citenamefont {Hook}, \citenamefont {Liu}, \citenamefont {Taylor},\ and\ \citenamefont {Zhao}}]{carney2019ultralight}%
  \BibitemOpen
  \bibfield  {author} {\bibinfo {author} {\bibfnamefont {D.}~\bibnamefont {Carney}}, \bibinfo {author} {\bibfnamefont {A.}~\bibnamefont {Hook}}, \bibinfo {author} {\bibfnamefont {Z.}~\bibnamefont {Liu}}, \bibinfo {author} {\bibfnamefont {J.~M.}\ \bibnamefont {Taylor}}, \ and\ \bibinfo {author} {\bibfnamefont {Y.}~\bibnamefont {Zhao}},\ }\bibfield  {title} {\enquote {\bibinfo {title} {Ultralight dark matter detection with mechanical quantum sensors},}\ }\href {https://iopscience.iop.org/article/10.1088/1367-2630/abd9e7/meta} {\bibfield  {journal} {\bibinfo  {journal} {New J. Phys.}\ }\textbf {\bibinfo {volume} {23}},\ \bibinfo {pages} {023041} (\bibinfo {year} {2021}{\natexlab{b}})}\BibitemShut {NoStop}%
\bibitem [{\citenamefont {Dey~Chowdhury}\ \emph {et~al.}(2026)\citenamefont {Dey~Chowdhury}, \citenamefont {Manley}, \citenamefont {Condos}, \citenamefont {Agrawal},\ and\ \citenamefont {Wilson}}]{chowdhury2025optomechanical}%
  \BibitemOpen
  \bibfield  {author} {\bibinfo {author} {\bibfnamefont {M.}~\bibnamefont {Dey~Chowdhury}}, \bibinfo {author} {\bibfnamefont {J.}~\bibnamefont {Manley}}, \bibinfo {author} {\bibfnamefont {C.}~\bibnamefont {Condos}}, \bibinfo {author} {\bibfnamefont {A.}~\bibnamefont {Agrawal}}, \ and\ \bibinfo {author} {\bibfnamefont {D.}~\bibnamefont {Wilson}},\ }\bibfield  {title} {\enquote {\bibinfo {title} {Optomechanical accelerometer search for ultralight dark matter},}\ }\href {https://link.aps.org/pdf/10.1103/16yd-qj5v} {\bibfield  {journal} {\bibinfo  {journal} {Physical Review D}\ }\textbf {\bibinfo {volume} {113}},\ \bibinfo {pages} {L121303} (\bibinfo {year} {2026})}\BibitemShut {NoStop}%
\bibitem [{\citenamefont {Xia}\ \emph {et~al.}(2023)\citenamefont {Xia}, \citenamefont {Agrawal}, \citenamefont {Pluchar}, \citenamefont {Brady}, \citenamefont {Liu}, \citenamefont {Zhuang}, \citenamefont {Wilson},\ and\ \citenamefont {Zhang}}]{xia2022entanglement}%
  \BibitemOpen
  \bibfield  {author} {\bibinfo {author} {\bibfnamefont {Y.}~\bibnamefont {Xia}}, \bibinfo {author} {\bibfnamefont {A.~R.}\ \bibnamefont {Agrawal}}, \bibinfo {author} {\bibfnamefont {C.~M.}\ \bibnamefont {Pluchar}}, \bibinfo {author} {\bibfnamefont {A.~J.}\ \bibnamefont {Brady}}, \bibinfo {author} {\bibfnamefont {Z.}~\bibnamefont {Liu}}, \bibinfo {author} {\bibfnamefont {Q.}~\bibnamefont {Zhuang}}, \bibinfo {author} {\bibfnamefont {D.~J.}\ \bibnamefont {Wilson}}, \ and\ \bibinfo {author} {\bibfnamefont {Z.}~\bibnamefont {Zhang}},\ }\bibfield  {title} {\enquote {\bibinfo {title} {Entanglement-enhanced optomechanical sensing},}\ }\href {https://www.nature.com/articles/s41566-023-01178-0.pdf} {\bibfield  {journal} {\bibinfo  {journal} {Nature Photonics}\ }\textbf {\bibinfo {volume} {17}},\ \bibinfo {pages} {470--477} (\bibinfo {year} {2023})}\BibitemShut {NoStop}%
\bibitem [{\citenamefont {Li}\ \emph {et~al.}(2026)\citenamefont {Li}, \citenamefont {Li}, \citenamefont {Wang}, \citenamefont {Wang}, \citenamefont {Tian}, \citenamefont {Shi},\ and\ \citenamefont {Zheng}}]{li2026quantum}%
  \BibitemOpen
  \bibfield  {author} {\bibinfo {author} {\bibfnamefont {Q.}~\bibnamefont {Li}}, \bibinfo {author} {\bibfnamefont {W.}~\bibnamefont {Li}}, \bibinfo {author} {\bibfnamefont {Y.}~\bibnamefont {Wang}}, \bibinfo {author} {\bibfnamefont {Y.}~\bibnamefont {Wang}}, \bibinfo {author} {\bibfnamefont {L.}~\bibnamefont {Tian}}, \bibinfo {author} {\bibfnamefont {S.}~\bibnamefont {Shi}}, \ and\ \bibinfo {author} {\bibfnamefont {Y.}~\bibnamefont {Zheng}},\ }\bibfield  {title} {\enquote {\bibinfo {title} {Quantum-enhanced optomechanical sensor network},}\ }\href {https://onlinelibrary.wiley.com/doi/abs/10.1002/lpor.202501636} {\bibfield  {journal} {\bibinfo  {journal} {Laser \& Photonics Reviews}\ }\textbf {\bibinfo {volume} {20}},\ \bibinfo {pages} {e01636} (\bibinfo {year} {2026})}\BibitemShut {NoStop}%
\bibitem [{\citenamefont {Brady}\ \emph {et~al.}(2023)\citenamefont {Brady}, \citenamefont {Chen}, \citenamefont {Xia}, \citenamefont {Manley}, \citenamefont {Dey~Chowdhury}, \citenamefont {Xiao}, \citenamefont {Liu}, \citenamefont {Harnik}, \citenamefont {Wilson}, \citenamefont {Zhang},\ and\ \citenamefont {Zhuang}}]{brady2023entanglement}%
  \BibitemOpen
  \bibfield  {author} {\bibinfo {author} {\bibfnamefont {A.~J.}\ \bibnamefont {Brady}}, \bibinfo {author} {\bibfnamefont {X.}~\bibnamefont {Chen}}, \bibinfo {author} {\bibfnamefont {Y.}~\bibnamefont {Xia}}, \bibinfo {author} {\bibfnamefont {J.}~\bibnamefont {Manley}}, \bibinfo {author} {\bibfnamefont {M.}~\bibnamefont {Dey~Chowdhury}}, \bibinfo {author} {\bibfnamefont {K.}~\bibnamefont {Xiao}}, \bibinfo {author} {\bibfnamefont {Z.}~\bibnamefont {Liu}}, \bibinfo {author} {\bibfnamefont {R.}~\bibnamefont {Harnik}}, \bibinfo {author} {\bibfnamefont {D.~J.}\ \bibnamefont {Wilson}}, \bibinfo {author} {\bibfnamefont {Z.}~\bibnamefont {Zhang}}, \ and\ \bibinfo {author} {\bibfnamefont {Q.}~\bibnamefont {Zhuang}},\ }\bibfield  {title} {\enquote {\bibinfo {title} {Entanglement-enhanced optomechanical sensor array with application to dark matter searches},}\ }\href {\doibase 10.1038/s42005-023-01357-z} {\bibfield  {journal} {\bibinfo  {journal} {Commun. Phys.}\ }\textbf {\bibinfo {volume} {6}},\ \bibinfo {pages} {237}
  (\bibinfo {year} {2023})}\BibitemShut {NoStop}%
\bibitem [{\citenamefont {Bereyhi}\ \emph {et~al.}(2019)\citenamefont {Bereyhi}, \citenamefont {Beccari}, \citenamefont {Fedorov}, \citenamefont {Ghadimi}, \citenamefont {Schilling}, \citenamefont {Wilson}, \citenamefont {Engelsen},\ and\ \citenamefont {Kippenberg}}]{bereyhi2019clamp}%
  \BibitemOpen
  \bibfield  {author} {\bibinfo {author} {\bibfnamefont {M.~J.}\ \bibnamefont {Bereyhi}}, \bibinfo {author} {\bibfnamefont {A.}~\bibnamefont {Beccari}}, \bibinfo {author} {\bibfnamefont {S.~A.}\ \bibnamefont {Fedorov}}, \bibinfo {author} {\bibfnamefont {A.~H.}\ \bibnamefont {Ghadimi}}, \bibinfo {author} {\bibfnamefont {R.}~\bibnamefont {Schilling}}, \bibinfo {author} {\bibfnamefont {D.~J.}\ \bibnamefont {Wilson}}, \bibinfo {author} {\bibfnamefont {N.~J.}\ \bibnamefont {Engelsen}}, \ and\ \bibinfo {author} {\bibfnamefont {T.~J.}\ \bibnamefont {Kippenberg}},\ }\bibfield  {title} {\enquote {\bibinfo {title} {Clamp-tapering increases the quality factor of stressed nanobeams},}\ }\href {https://pubs.acs.org/doi/abs/10.1021/acs.nanolett.8b04942} {\bibfield  {journal} {\bibinfo  {journal} {Nano letters}\ }\textbf {\bibinfo {volume} {19}},\ \bibinfo {pages} {2329--2333} (\bibinfo {year} {2019})}\BibitemShut {NoStop}%
\bibitem [{\citenamefont {Villanueva}\ and\ \citenamefont {Schmid}(2014)}]{villanueva2014evidence}%
  \BibitemOpen
  \bibfield  {author} {\bibinfo {author} {\bibfnamefont {L.~G.}\ \bibnamefont {Villanueva}}\ and\ \bibinfo {author} {\bibfnamefont {S.}~\bibnamefont {Schmid}},\ }\bibfield  {title} {\enquote {\bibinfo {title} {Evidence of surface loss as ubiquitous limiting damping mechanism in sin micro-and nanomechanical resonators},}\ }\href {https://doi.org/10.1103/PhysRevLett.113.227201} {\bibfield  {journal} {\bibinfo  {journal} {Physical Review Letters}\ }\textbf {\bibinfo {volume} {113}},\ \bibinfo {pages} {227201} (\bibinfo {year} {2014})}\BibitemShut {NoStop}%
\bibitem [{tet()}]{tetherWidth}%
  \BibitemOpen
  \href@noop {} {}\bibinfo {note} {We optimized sail membranes over a range of tether widths and only saw significant changes in the frequency between each design---Q-m product remained roughly constant. We opted to focus on devices with 10 and 15 micron tethers because of their predicted performance and robustness during fabrication.}\BibitemShut {Stop}%
\bibitem [{\citenamefont {Sadeghi}\ \emph {et~al.}(2019)\citenamefont {Sadeghi}, \citenamefont {Tanzer}, \citenamefont {Christensen},\ and\ \citenamefont {Schmid}}]{sadeghi2019influence}%
  \BibitemOpen
  \bibfield  {author} {\bibinfo {author} {\bibfnamefont {P.}~\bibnamefont {Sadeghi}}, \bibinfo {author} {\bibfnamefont {M.}~\bibnamefont {Tanzer}}, \bibinfo {author} {\bibfnamefont {S.~L.}\ \bibnamefont {Christensen}}, \ and\ \bibinfo {author} {\bibfnamefont {S.}~\bibnamefont {Schmid}},\ }\bibfield  {title} {\enquote {\bibinfo {title} {Influence of clamp-widening on the quality factor of nanomechanical silicon nitride resonators},}\ }\href {https://doi.org/10.1063/1.5111712} {\bibfield  {journal} {\bibinfo  {journal} {Journal of Applied Physics}\ }\textbf {\bibinfo {volume} {126}} (\bibinfo {year} {2019})}\BibitemShut {NoStop}%
\bibitem [{\citenamefont {Norder}\ \emph {et~al.}(2026)\citenamefont {Norder}, \citenamefont {Ke{\c{s}}kekler},\ and\ \citenamefont {Norte}}]{norder2026high}%
  \BibitemOpen
  \bibfield  {author} {\bibinfo {author} {\bibfnamefont {L.}~\bibnamefont {Norder}}, \bibinfo {author} {\bibfnamefont {A.}~\bibnamefont {Ke{\c{s}}kekler}}, \ and\ \bibinfo {author} {\bibfnamefont {R.~A.}\ \bibnamefont {Norte}},\ }\bibfield  {title} {\enquote {\bibinfo {title} {High-power laser drives motion in ultra-thin photonic crystal lightsails via radiation pressure},}\ }\href {https://arxiv.org/abs/2606.20149} {\bibfield  {journal} {\bibinfo  {journal} {arXiv preprint arXiv:2606.20149}\ } (\bibinfo {year} {2026})}\BibitemShut {NoStop}%
\bibitem [{\citenamefont {Hyatt}\ \emph {et~al.}(2025{\natexlab{b}})\citenamefont {Hyatt}, \citenamefont {Flores}, \citenamefont {Agrawal}, \citenamefont {Condos},\ and\ \citenamefont {Wilson}}]{hyatt2025fabrication}%
  \BibitemOpen
  \bibfield  {author} {\bibinfo {author} {\bibfnamefont {A.~D.}\ \bibnamefont {Hyatt}}, \bibinfo {author} {\bibfnamefont {O.~A.}\ \bibnamefont {Flores}}, \bibinfo {author} {\bibfnamefont {A.~R.}\ \bibnamefont {Agrawal}}, \bibinfo {author} {\bibfnamefont {C.~A.}\ \bibnamefont {Condos}}, \ and\ \bibinfo {author} {\bibfnamefont {D.~J.}\ \bibnamefont {Wilson}},\ }\bibfield  {title} {\enquote {\bibinfo {title} {Fabrication and characterization of high-q silicon nitride membrane resonators},}\ }\href {\doibase https://app.jove.com/v/68706/fabrication-characterization-high-q-silicon-nitride-membrane} {\bibfield  {journal} {\bibinfo  {journal} {JoVE}\ ,\ \bibinfo {pages} {e68706}} (\bibinfo {year} {2025}{\natexlab{b}})}\BibitemShut {NoStop}%
\bibitem [{\citenamefont {Wilson}\ \emph {et~al.}(2009)\citenamefont {Wilson}, \citenamefont {Regal}, \citenamefont {Papp},\ and\ \citenamefont {Kimble}}]{wilson2009cavity}%
  \BibitemOpen
  \bibfield  {author} {\bibinfo {author} {\bibfnamefont {D.~J.}\ \bibnamefont {Wilson}}, \bibinfo {author} {\bibfnamefont {C.~A.}\ \bibnamefont {Regal}}, \bibinfo {author} {\bibfnamefont {S.~B.}\ \bibnamefont {Papp}}, \ and\ \bibinfo {author} {\bibfnamefont {H.}~\bibnamefont {Kimble}},\ }\bibfield  {title} {\enquote {\bibinfo {title} {Cavity optomechanics with stoichiometric sin films},}\ }\href {https://journals.aps.org/prl/abstract/10.1103/PhysRevLett.103.207204} {\bibfield  {journal} {\bibinfo  {journal} {Physical Review Letters}\ }\textbf {\bibinfo {volume} {103}},\ \bibinfo {pages} {207204} (\bibinfo {year} {2009})}\BibitemShut {NoStop}%
\bibitem [{\citenamefont {Pluchar}\ \emph {et~al.}(2025)\citenamefont {Pluchar}, \citenamefont {Agrawal},\ and\ \citenamefont {Wilson}}]{pluchar2025quantum}%
  \BibitemOpen
  \bibfield  {author} {\bibinfo {author} {\bibfnamefont {C.~M.}\ \bibnamefont {Pluchar}}, \bibinfo {author} {\bibfnamefont {A.~R.}\ \bibnamefont {Agrawal}}, \ and\ \bibinfo {author} {\bibfnamefont {D.~J.}\ \bibnamefont {Wilson}},\ }\bibfield  {title} {\enquote {\bibinfo {title} {Quantum-limited optical lever measurement of a torsion oscillator},}\ }\href {https://opg.optica.org/optica/abstract.cfm?URI=optica-12-3-418} {\bibfield  {journal} {\bibinfo  {journal} {Optica}\ }\textbf {\bibinfo {volume} {12}},\ \bibinfo {pages} {418--423} (\bibinfo {year} {2025})}\BibitemShut {NoStop}%
\bibitem [{\citenamefont {Green}\ \emph {et~al.}(2025)\citenamefont {Green}, \citenamefont {Bao}, \citenamefont {Lawall}, \citenamefont {Gorman},\ and\ \citenamefont {Barker}}]{green2025accurate}%
  \BibitemOpen
  \bibfield  {author} {\bibinfo {author} {\bibfnamefont {O.~R.}\ \bibnamefont {Green}}, \bibinfo {author} {\bibfnamefont {Y.}~\bibnamefont {Bao}}, \bibinfo {author} {\bibfnamefont {J.~R.}\ \bibnamefont {Lawall}}, \bibinfo {author} {\bibfnamefont {J.~J.}\ \bibnamefont {Gorman}}, \ and\ \bibinfo {author} {\bibfnamefont {D.~S.}\ \bibnamefont {Barker}},\ }\bibfield  {title} {\enquote {\bibinfo {title} {Accurate, precise pressure sensing with tethered optomechanics},}\ }\href {https://link.aps.org/pdf/10.1103/9dtb-sk2j} {\bibfield  {journal} {\bibinfo  {journal} {Physical Review Applied}\ }\textbf {\bibinfo {volume} {24}},\ \bibinfo {pages} {024069} (\bibinfo {year} {2025})}\BibitemShut {NoStop}%
\bibitem [{\citenamefont {Reinhardt}\ \emph {et~al.}(2024)\citenamefont {Reinhardt}, \citenamefont {Masalehdan}, \citenamefont {Croatto}, \citenamefont {Franke}, \citenamefont {Kunze}, \citenamefont {Schaffran}, \citenamefont {Sueltmann}, \citenamefont {Lindner},\ and\ \citenamefont {Schnabel}}]{reinhardt2024self}%
  \BibitemOpen
  \bibfield  {author} {\bibinfo {author} {\bibfnamefont {C.}~\bibnamefont {Reinhardt}}, \bibinfo {author} {\bibfnamefont {H.}~\bibnamefont {Masalehdan}}, \bibinfo {author} {\bibfnamefont {S.}~\bibnamefont {Croatto}}, \bibinfo {author} {\bibfnamefont {A.}~\bibnamefont {Franke}}, \bibinfo {author} {\bibfnamefont {M.~B.}\ \bibnamefont {Kunze}}, \bibinfo {author} {\bibfnamefont {J.}~\bibnamefont {Schaffran}}, \bibinfo {author} {\bibfnamefont {N.}~\bibnamefont {Sueltmann}}, \bibinfo {author} {\bibfnamefont {A.}~\bibnamefont {Lindner}}, \ and\ \bibinfo {author} {\bibfnamefont {R.}~\bibnamefont {Schnabel}},\ }\bibfield  {title} {\enquote {\bibinfo {title} {Self-calibrating gas pressure sensor with a 10-decade measurement range},}\ }\href {https://pubs.acs.org/doi/pdf/10.1021/acsphotonics.3c01488} {\bibfield  {journal} {\bibinfo  {journal} {ACS photonics}\ }\textbf {\bibinfo {volume} {11}},\ \bibinfo {pages} {1438--1446} (\bibinfo {year} {2024})}\BibitemShut {NoStop}%
\bibitem [{\citenamefont {de~Jong}\ \emph {et~al.}(2022)\citenamefont {de~Jong}, \citenamefont {ten Wolde}, \citenamefont {Cupertino}, \citenamefont {Gr{\"o}blacher}, \citenamefont {Steeneken},\ and\ \citenamefont {Norte}}]{de2022mechanical}%
  \BibitemOpen
  \bibfield  {author} {\bibinfo {author} {\bibfnamefont {M.~H.}\ \bibnamefont {de~Jong}}, \bibinfo {author} {\bibfnamefont {M.~A.}\ \bibnamefont {ten Wolde}}, \bibinfo {author} {\bibfnamefont {A.}~\bibnamefont {Cupertino}}, \bibinfo {author} {\bibfnamefont {S.}~\bibnamefont {Gr{\"o}blacher}}, \bibinfo {author} {\bibfnamefont {P.~G.}\ \bibnamefont {Steeneken}}, \ and\ \bibinfo {author} {\bibfnamefont {R.~A.}\ \bibnamefont {Norte}},\ }\bibfield  {title} {\enquote {\bibinfo {title} {Mechanical dissipation by substrate--mode coupling in sin resonators},}\ }\href {https://doi.org/10.1063/5.0092894} {\bibfield  {journal} {\bibinfo  {journal} {Applied Physics Letters}\ }\textbf {\bibinfo {volume} {121}} (\bibinfo {year} {2022})}\BibitemShut {NoStop}%
\bibitem [{Pol()}]{Polytec}%
  \BibitemOpen
  \href@noop {} {}\bibinfo {note} {Polytec VibroScan-QTec.}\BibitemShut {Stop}%
\bibitem [{bal()}]{balancedDetCav}%
  \BibitemOpen
  \href@noop {} {}\bibinfo {note} {The laser beam was split before passing through the cavity, creating an auxiliary reference beam. The transmitted and referenced beams were combined on a balanced photodetector to cancel classical (technical) laser noise. This results in an extra factor of two in our expression for intensity shot noise, $S_P^\t{imp}$.}\BibitemShut {Stop}%
\bibitem [{\citenamefont {Michaeli}\ \emph {et~al.}(2025{\natexlab{a}})\citenamefont {Michaeli}, \citenamefont {Gao}, \citenamefont {Kelzenberg}, \citenamefont {Hail}, \citenamefont {Merkt}, \citenamefont {Sader},\ and\ \citenamefont {Atwater}}]{michaeli2025direct}%
  \BibitemOpen
  \bibfield  {author} {\bibinfo {author} {\bibfnamefont {L.}~\bibnamefont {Michaeli}}, \bibinfo {author} {\bibfnamefont {R.}~\bibnamefont {Gao}}, \bibinfo {author} {\bibfnamefont {M.~D.}\ \bibnamefont {Kelzenberg}}, \bibinfo {author} {\bibfnamefont {C.~U.}\ \bibnamefont {Hail}}, \bibinfo {author} {\bibfnamefont {A.}~\bibnamefont {Merkt}}, \bibinfo {author} {\bibfnamefont {J.~E.}\ \bibnamefont {Sader}}, \ and\ \bibinfo {author} {\bibfnamefont {H.~A.}\ \bibnamefont {Atwater}},\ }\bibfield  {title} {\enquote {\bibinfo {title} {Direct radiation pressure measurements for lightsail membranes},}\ }\href {https://www.nature.com/articles/s41566-024-01605-w} {\bibfield  {journal} {\bibinfo  {journal} {Nature Photonics}\ }\textbf {\bibinfo {volume} {19}},\ \bibinfo {pages} {369--377} (\bibinfo {year} {2025}{\natexlab{a}})}\BibitemShut {NoStop}%
\bibitem [{\citenamefont {Michaeli}\ \emph {et~al.}(2025{\natexlab{b}})\citenamefont {Michaeli}, \citenamefont {Gao}, \citenamefont {Kelzenberg}, \citenamefont {Hail}, \citenamefont {Sader},\ and\ \citenamefont {Atwater}}]{michaeli2025optically}%
  \BibitemOpen
  \bibfield  {author} {\bibinfo {author} {\bibfnamefont {L.}~\bibnamefont {Michaeli}}, \bibinfo {author} {\bibfnamefont {R.}~\bibnamefont {Gao}}, \bibinfo {author} {\bibfnamefont {M.~D.}\ \bibnamefont {Kelzenberg}}, \bibinfo {author} {\bibfnamefont {C.~U.}\ \bibnamefont {Hail}}, \bibinfo {author} {\bibfnamefont {J.~E.}\ \bibnamefont {Sader}}, \ and\ \bibinfo {author} {\bibfnamefont {H.~A.}\ \bibnamefont {Atwater}},\ }\bibfield  {title} {\enquote {\bibinfo {title} {Optically actuated transitions in multimodal, bistable micromechanical oscillators},}\ }\href {https://arxiv.org/abs/2507.22605} {\bibfield  {journal} {\bibinfo  {journal} {arXiv preprint arXiv:2507.22605}\ } (\bibinfo {year} {2025}{\natexlab{b}})}\BibitemShut {NoStop}%
\bibitem [{\citenamefont {Zhou}\ \emph {et~al.}(2021)\citenamefont {Zhou}, \citenamefont {Bao}, \citenamefont {Madugani}, \citenamefont {Long}, \citenamefont {Gorman},\ and\ \citenamefont {LeBrun}}]{zhou2021broadband}%
  \BibitemOpen
  \bibfield  {author} {\bibinfo {author} {\bibfnamefont {F.}~\bibnamefont {Zhou}}, \bibinfo {author} {\bibfnamefont {Y.}~\bibnamefont {Bao}}, \bibinfo {author} {\bibfnamefont {R.}~\bibnamefont {Madugani}}, \bibinfo {author} {\bibfnamefont {D.~A.}\ \bibnamefont {Long}}, \bibinfo {author} {\bibfnamefont {J.~J.}\ \bibnamefont {Gorman}}, \ and\ \bibinfo {author} {\bibfnamefont {T.~W.}\ \bibnamefont {LeBrun}},\ }\bibfield  {title} {\enquote {\bibinfo {title} {Broadband thermomechanically limited sensing with an optomechanical accelerometer},}\ }\href {https://opg.optica.org/viewmedia.cfm?uri=optica-8-3-350&seq=0} {\bibfield  {journal} {\bibinfo  {journal} {Optica}\ }\textbf {\bibinfo {volume} {8}},\ \bibinfo {pages} {350--356} (\bibinfo {year} {2021})}\BibitemShut {NoStop}%
\bibitem [{\citenamefont {St-Gelais}\ \emph {et~al.}(2019)\citenamefont {St-Gelais}, \citenamefont {Bernard}, \citenamefont {Reinhardt},\ and\ \citenamefont {Sankey}}]{st2019swept}%
  \BibitemOpen
  \bibfield  {author} {\bibinfo {author} {\bibfnamefont {R.}~\bibnamefont {St-Gelais}}, \bibinfo {author} {\bibfnamefont {S.}~\bibnamefont {Bernard}}, \bibinfo {author} {\bibfnamefont {C.}~\bibnamefont {Reinhardt}}, \ and\ \bibinfo {author} {\bibfnamefont {J.~C.}\ \bibnamefont {Sankey}},\ }\bibfield  {title} {\enquote {\bibinfo {title} {Swept-frequency drumhead optomechanical resonators},}\ }\href {https://pubs.acs.org/doi/full/10.1021/acsphotonics.8b01519} {\bibfield  {journal} {\bibinfo  {journal} {ACS Photonics}\ }\textbf {\bibinfo {volume} {6}},\ \bibinfo {pages} {525--530} (\bibinfo {year} {2019})}\BibitemShut {NoStop}%
\bibitem [{\citenamefont {Agrawal}\ \emph {et~al.}(2024)\citenamefont {Agrawal}, \citenamefont {Manley}, \citenamefont {Allepuz-Requena},\ and\ \citenamefont {Wilson}}]{agrawal2024focusing}%
  \BibitemOpen
  \bibfield  {author} {\bibinfo {author} {\bibfnamefont {A.}~\bibnamefont {Agrawal}}, \bibinfo {author} {\bibfnamefont {J.}~\bibnamefont {Manley}}, \bibinfo {author} {\bibfnamefont {D.}~\bibnamefont {Allepuz-Requena}}, \ and\ \bibinfo {author} {\bibfnamefont {D.}~\bibnamefont {Wilson}},\ }\bibfield  {title} {\enquote {\bibinfo {title} {Focusing membrane metamirrors for integrated cavity optomechanics},}\ }\href {\doibase 10.1364/OPTICA.522509} {\bibfield  {journal} {\bibinfo  {journal} {Optica}\ }\textbf {\bibinfo {volume} {11}},\ \bibinfo {pages} {1235--1241} (\bibinfo {year} {2024})}\BibitemShut {NoStop}%
\end{thebibliography}%

\end{document}